\documentclass[pdflatex,sn-nature]{sn-jnl}

\usepackage{graphicx}%
\usepackage{multirow}%
\usepackage{amsmath,amssymb,amsfonts}%
\usepackage{amsthm}%
\usepackage{mathrsfs}%
\usepackage[title]{appendix}%
\usepackage{xcolor}%
\usepackage{textcomp}%
\usepackage{manyfoot}%
\usepackage{booktabs}%
\usepackage{algorithm}%
\usepackage{algorithmicx}%
\usepackage{algpseudocode}%
\usepackage{listings}%
\usepackage{float}
\usepackage{rotating}
\usepackage{makecell}
\usepackage{multirow}
\usepackage{array}
\usepackage[table]{xcolor} 
\usepackage[version=4]{mhchem}
\usepackage{subcaption}
\usepackage[normalem]{ulem}
\usepackage{tabularx}
\usepackage{adjustbox}

\usepackage{xcolor}
\usepackage{colortbl}
\usepackage{pgfmath}
\newlength{\BaseColWidth}
\newcolumntype{L}{>{\raggedright\arraybackslash}p{0.3\BaseColWidth}}
\newcolumntype{D}{>{\raggedright\arraybackslash}p{0.4\BaseColWidth}}
\newcolumntype{V}{>{\raggedright\arraybackslash}p{0.3\BaseColWidth}}

\theoremstyle{thmstyleone}%
\theoremstyle{thmstyletwo}%

\theoremstyle{thmstylethree}%

\newcommand{\CellMin}{0}
\newcommand{\CellMax}{0.4}
\colorlet{cmin}{green!10!white}  
\colorlet{cmed}{yellow!20!white} 
\colorlet{cmax}{red!20!white}    
\newcommand{\ColorCell}[1]{%
  \begingroup
  \pgfmathsetmacro{\val}{#1}%
  \pgfmathsetmacro{\norm}{(\val-\CellMin)/(\CellMax-\CellMin)}%
  \pgfmathsetmacro{\norm}{max(min(\norm,1),0)}%
  \pgfmathsetmacro{\pct}{\norm*100}%
  \pgfmathtruncatemacro{\pctint}{round(\pct)}%

  \ifnum\pctint<50
    \pgfmathtruncatemacro{\mix}{round(100 - 2*\pctint)}%
    \edef\temp{\noexpand\cellcolor{cmin!\mix!cmed}}%
  \else
    \pgfmathtruncatemacro{\mix}{round(200 - 2*\pctint)}%
    \edef\temp{\noexpand\cellcolor{cmed!\mix!cmax}}%
  \fi
  \temp #1%
  \endgroup
}

\newcommand{\ColorCellMM}[3]{%
  \begingroup
  \renewcommand{\CellMin}{#1}%
  \renewcommand{\CellMax}{#2}%
  \ColorCell{#3}%
  \endgroup
}

\newcommand{\ColorCellBG}[1]{%
  \begingroup
  \pgfmathsetmacro{\val}{#1}%
  \pgfmathsetmacro{\norm}{(\val-\CellMin)/(\CellMax-\CellMin)}%
  \pgfmathsetmacro{\norm}{max(min(\norm,1),0)}%
  \pgfmathsetmacro{\pct}{\norm*100}%
  \pgfmathtruncatemacro{\pctint}{round(\pct)}%
  \ifnum\pctint<50
    \pgfmathtruncatemacro{\mix}{round(100 - 2*\pctint)}%
    \edef\temp{\noexpand\cellcolor{cmin!\mix!cmed}}%
  \else
    \pgfmathtruncatemacro{\mix}{round(200 - 2*\pctint)}%
    \edef\temp{\noexpand\cellcolor{cmed!\mix!cmax}}%
  \fi
  \temp
  \endgroup
}

\newcommand{\ColorCellCSMM}[3]{%
  \begingroup
  \pgfmathsetmacro{\range}{#2-#1}%
  \renewcommand{\CellMin}{0}%
  \renewcommand{\CellMax}{\range}%
  \pgfmathsetmacro{\bad}{#2-#3}    
  \ColorCellBG{\bad}
  #3
  \endgroup
}

\title{Scalable Machine Learning Force Fields for Macromolecular Systems Through Long-Range Aware Message Passing}
\author[1]{\fnm{Chu} \sur{Wang}}
\equalcont{These authors contributed equally to this work. Sort by random order.}

\author[1]{\fnm{Lin} \sur{Huang}}
\equalcont{These authors contributed equally to this work. Sort by random order.}

\author[2]{\fnm{Xinran} \sur{Wei}}
\equalcont{These authors contributed equally to this work. Sort by random order.}

\author[2]{\fnm{Tao} \sur{Qin}}
\author[1]{\fnm{Arthur} \sur{Jiang}}
\author*[3]{\fnm{Lixue} \sur{Cheng}} \email{lixuecheng@ust.hk}
\author*[1]{\fnm{Jia} \sur{Zhang}}\email{jialrs.z@iquestlab.com}

\affil[1]{\orgdiv{UBio Team}, \orgname{IQuest Research}, \orgaddress{\street{No. 1 East Zhongguancun Road}, \city{Beijing}, \postcode{100084}, \country{China}}}
\affil[2]{\orgdiv{Zhongguancun Academy}, \orgname{Zhongguancun Institute of Artificial Intelligence}, \orgaddress{\street{No. 17 Daniufang Road}, \city{Beijing}, \postcode{100094}, \country{China}}}
\affil[3]{\orgdiv{Department of Chemistry}, \orgname{The Hong Kong Science and Technology}, \orgaddress{Kowloon, Hong Kong 999077, China}}

\begin{document}
\maketitle

\begin{abstract}

Machine learning force fields (MLFFs) have revolutionized molecular simulations by providing quantum mechanical accuracy at the speed of molecular mechanical computations. However, a fundamental reliance of these models on fixed-cutoff architectures limits their applicability to macromolecular systems where long-range interactions dominate. We demonstrate that this locality constraint causes force prediction errors to scale monotonically with system size, revealing a critical architectural bottleneck. To overcome this, we establish the systematically designed \emph{MolLR25} (\textit{Mol}ecules with \textit{L}ong-\textit{R}ange effect) benchmark up to 1200 atoms, generated using high-fidelity DFT, and introduce \emph{E2Former-LSR}, an equivariant transformer that explicitly integrates long-range attention blocks. E2Former-LSR exhibits stable error scaling, achieves superior fidelity in capturing non-covalent decay, and maintains precision on complex protein conformations. Crucially, its efficient design provides up to 30\% speedup compared to purely local models. This work validates the necessity of non-local architectures for generalizable MLFFs, enabling high-fidelity molecular dynamics for large-scale chemical and biological systems.

\end{abstract}

\section*{Introduction}
Machine learning has emerged as a transformative technology in molecular modeling, enabling simulations with quantum-level accuracy at a fraction of the computational cost. Among the most impactful developments,  \textit{machine learning force fields} (MLFFs) stand out by learning to approximate the potential energy surface and interatomic forces from high-level quantum mechanical data. These models now play a central role in the prediction of molecular properties, conformational sampling, molecular dynamics, and structure-based drug discovery.

A wave of early models—such as SchNet~\cite{schnet}, PhysNet~\cite{physnet}, and sGDML~\cite{sgdml}—demonstrated the feasibility of achieving DFT-level accuracy through neural networks by leveraging symmetry-aware architectures. The evolution of equivariant deep learning gave rise to models that explicitly respect spatial symmetries, including DeePMD~\cite{deepmd:wang2018deepmd,deepmd2:zeng2023deepmd,deepmd3:zeng2025deepmd}, NequIP~\cite{nequip}, PaiNN~\cite{painn}, and SpookyNet~\cite{spookynet}, which significantly improved sample efficiency and generalization to unseen molecules. More recent architectures—such as GemNet~\cite{gemnet}, Allegro~\cite{allegro}, TorchMD-Net~\cite{md22}, DPA~\cite{dpa:zhang2024pretraining,dpa2:zhang2024dpa}, Uni-Mol~\cite{uni-mol:zhouuni,uni-mol2:ji2024uni}—have advanced the frontier of MLFFs by incorporating physically informed message-passing, directional filters, and graph attention mechanisms, enabling scalable training on larger datasets and improving stability for long molecular dynamics simulations.
In parallel, models such as the Equiformer series~\cite{equiformer, liaoequiformerv2}, ViSNet~\cite{visnet}, MACE~\cite{mace}, SE(3)-Transformer~\cite{se3transfuchs2020se}, UMA~\cite{uma:wood2025family}, and SimPoly~\cite{simm2025simpoly} have focused on enhancing expressivity by capturing higher-order geometric features and nonlocal dependencies while retaining SE(3)/E(3)-equivariance. These innovations allow force fields to better model complex systems requiring quantum-level accuracy, such as non-covalent interactions and long-range polarization. Collectively, these MLFFs have laid a strong foundation for accurate simulations across a broad spectrum of molecules and materials.

Despite these remarkable developments, a critical limitation remains: most current MLFFs are designed and evaluated almost exclusively on \textit{small molecules}, typically containing fewer than 300 atoms. Benchmark datasets such as \textit{QM9}~\cite{qm9} and \textit{MD17}~\cite{md17} exemplify this regime, focusing on small organic molecules with limited topological diversity. The more recent \textit{MD22} dataset~\cite{md22} and \textit{OMol25} dataset~\cite{omol25:levine2025open} sought to broaden this scope, offering molecules with more than 300 atoms. While MD22 and OMol25 represent an important step toward larger-scale modeling, they still fall short of representing realistic macromolecular systems such as proteins, metal--organic frameworks (MOFs), or solvated complexes, which often span thousands of atoms and exhibit rich long-range interactions. The root of this limitation lies in the computational cost of reference data generation: conventional DFT methods scale as $\mathcal{O}\left(\mathcal{N}^3\right)$ to $\mathcal{O}\left(\mathcal{N}^4\right)$ with system size $\mathcal{N}$, making the generation of accurate quantum labels for large molecules computationally prohibitive. As a result, even the most advanced MLFFs rely on \textit{locality assumptions}, typically truncating interactions beyond a fixed cutoff radius to manage complexity. While effective in reducing costs and improving scalability, this \textit{local modeling paradigm} inherently neglects long-range interactions—such as dispersion forces, distant electrostatics, or through-space polarization—which become increasingly significant in large and complex molecular systems. While several prior efforts have attempted to explicitly model long-range interactions (e.g., \cite{lilong}), a systematic evaluation and rigorous validation remain critically underdeveloped due to the scarcity of relevant benchmark data.

\begin{figure}[p]
\centering
\includegraphics[width=1.0\linewidth,height=1.0\textheight,keepaspectratio]{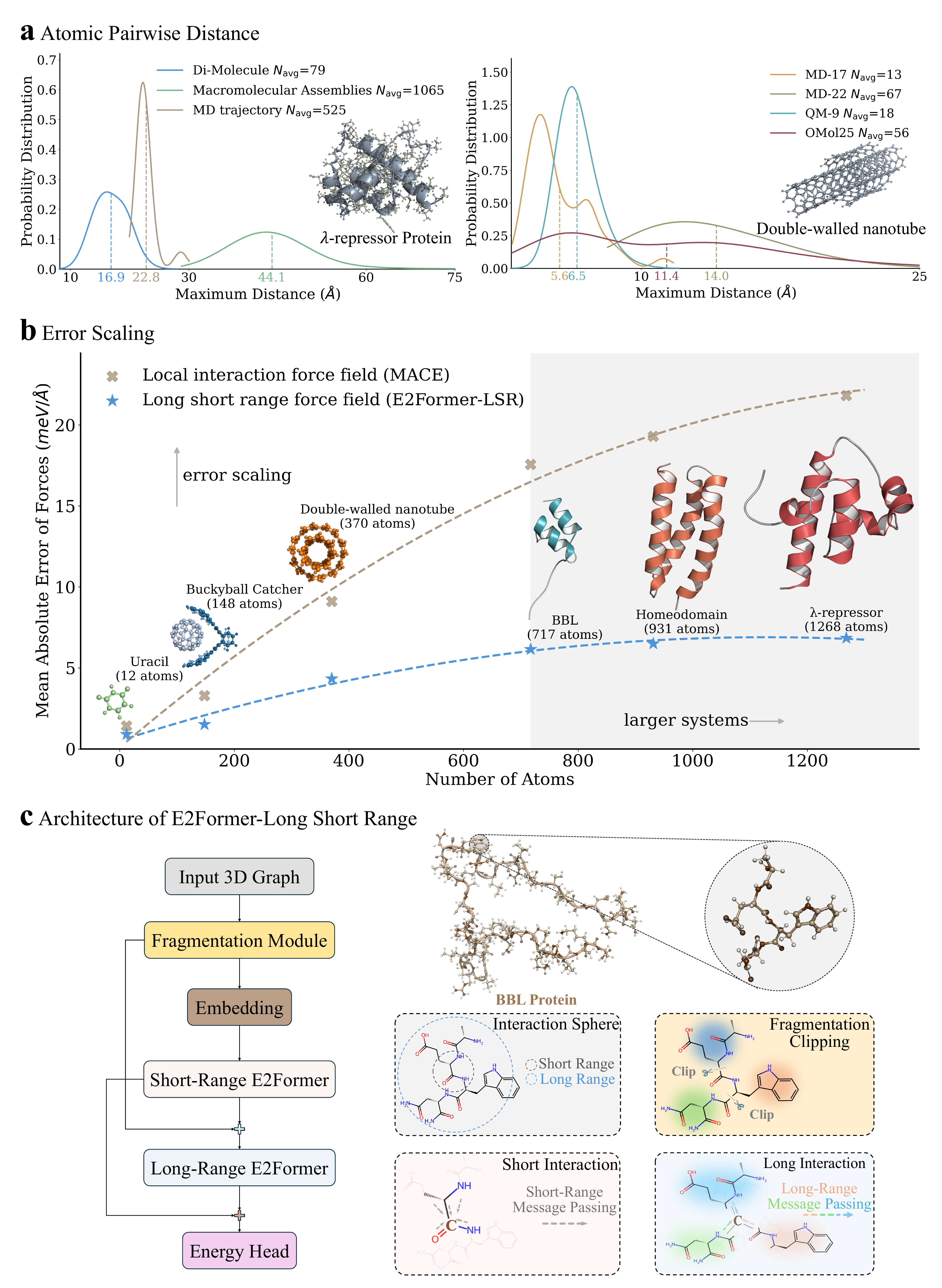}
\captionsetup{width=\linewidth}
\caption{\textbf{Architectural necessity and benchmark scope for long-range MLFFs. } 
\textbf{a}. Distribution of system size ($\mathcal{N}$) and maximum inter-atomic distance ($R_{\text{max}}$) for common benchmark datasets (QM9, MD17, MD22, OMol25) compared to the proposed $\emph{MolLR25}$ suite. The ${MolLR25}$ data extends significantly beyond existing benchmarks, covering systems up to $\mathcal{N} \approx 1200$ atoms and spatial ranges up to $R_{\text{max}} \approx 75\text{\r{A}}$. 
\textbf{b}. Scaling behavior of the training error ($\text{RMSE}_{\text{train}}$) versus system size ($\mathcal{N}$) for the local $\text{MACE-large}$ model and the non-local $\text{E2Former-LSR}$. $\text{E2Former-LSR}$ exhibits a notably flatter and converging error curve as complexity increases, demonstrating its robust capacity to integrate long-range dependencies. 
\textbf{c}. Schematic of the $\text{E2Former-LSR}$ architecture. The design overcomes the fixed-cutoff limit by segmenting the molecule into fragments and utilizing a transformer-based attention mechanism to jointly model short-range atom-atom interactions and long-range atom-fragment interactions for comprehensive feature aggregation.}

\label{fig:main:1}
\end{figure}

To rigorously quantify the limitation of pure local models in the context of large-scale systems, we constructed a bespoke set of DFT reference data for systems extending up to 1200 atoms, designed explicitly to assess long-range fidelity across three crucial regimes: isolated interactions, complex static environments, and dynamic stability, that is, the \emph{MolLR25} (molecule with long-range effect) dataset. Specifically, the MolLR25 includes: the Di-dataset dissociation dataset for testing smooth, asymptotic non-bonded energy and force decay over large distances; medium-scale protein assemblies derived from D.E. Shaw's MD trajectory data~\cite{deshaw:lindorff2011fast}, providing challenging, high-density environments for static evaluation on realistic proteins; and extended MD trajectory data for systems up to more than 500 atoms, essential for assessing the long-term fidelity and stability of the predicted potential energy surface in dynamic simulations. The DFT calculations were performed using the established def2-SVP basis set \cite{weigend2005balanced} and the highly accurate $\omega\text{B97X}$ double hybrid functional coupled with Grimme's D3 dispersion correction \cite{grimme2010consistent}; this methodological choice was crucial to ensure reliable quantum labels for the extended, non-covalent interactions inherent to large molecular assemblies. 
To visually confirm the enhanced scale and spatial complexity of our new benchmark suite, Figure \ref{fig:main:1}.a summarizes the average atom count ($\mathcal{N}_{\text{avg}}$) and the distribution of the maximum inter-atomic distance ($R_{\text{max}}$) for both existing datasets and our three proposed datasets. This visualization clearly demonstrates that our newly constructed data significantly exceed the size ($\mathcal{N}$) and spatial extent ($R_{\text{max}}$) of existing benchmarks. For instance, the maximum average atom count in \emph{MolLR25} reaches $\mathcal{N}_{\text{avg}} = 1065$, compared to the largest existing average of 67 atoms in MD22. Similarly, our maximum $R_{\text{max}}$ extends up to $75\text{\r{A}}$, substantially surpassing the current limit of less than $30\text{\r{A}}$. This scale ensures the rigorous assessment of models in the regime where long-range effects dominate.

We first assessed the intrinsic fitting capacity of a leading local model, MACE-large~\cite{mace-off:kovacs2025mace}, across this broad range of molecular sizes. As depicted in Figure \ref{fig:main:1}.b, 
while MACE achieves remarkable fidelity on numerous smaller benchmarks, its mean absolute force prediction error on the training data is observed to increase systematically with system size ($\mathcal{N}$). 
This stark degradation confirms that only using local information limits the model's ability to integrate necessary long-range dependencies across the system's growing complexity. This failure persists even when the target quantum labels are available during training, underscoring a fundamental architectural constraint of local modeling paradigms. 

To address this, we introduce E2Former-LSR, which integrates the Long-Short-Range (LSR) message passing framework \cite{lilong} with the state-of-the-art E2Former architecture \cite{e2former}. This design represents a fundamental departure from the conventional fixed-cutoff paradigm by explicitly and jointly modeling both short-range and long-range interactions in a unified framework. As shown in Figure \ref{fig:main:1}.c, E2Former-LSR is a transformer-like equivariant neural network and employs an alternating block design, where local message-passing layers capture fine-grained covalent bonding and strong repulsion, while dedicated distant attention blocks dynamically aggregate features across spatially separated atomic clusters. As evidenced in Figure \ref{fig:main:1}.b, E2Former-LSR exhibits a dramatically different error scaling: its force prediction error remains nearly constant and low, even as the system size extends beyond 1200 atoms. This remarkable stability demonstrates the model's ability to systematically learn the full quantum mechanical interaction spectrum, enabling reliable and accurate macromolecular modeling beyond 1000 atoms, a regime where purely local methods inherently fail. Crucially, due to its efficient transformer-like architectural design, E2Former-LSR achieves a significant computational advantage: despite explicitly modeling long-range interactions, it provides a speedup of up to 30\% compared to purely local models, such as MACE (Table \ref{tab:2:deshaw} in Supplementary Information).

The observed error scaling in Figure \ref{fig:main:1}.b provides an initial compelling demonstration of the superiority of explicitly modeling long-range interactions. We now proceed with a more rigorous and detailed empirical validation of E2Former-LSR's fidelity across the three crucial regimes defined by our comprehensive benchmark suite: the long-range interaction dissociation test, the medium-scale protein conformation fidelity test, and the large-molecule MD trajectory stability test.

\section*{Results}
\subsection*{Overview of E2Former-LSR Architecture}
To systematically overcome the locality limitations demonstrated above, we developed the E2Former-LSR architecture, leveraging and extending the foundational E2Former model \cite{e2former}. The core E2Former framework is a transformer-like equivariant neural network that utilizes self-attention mechanisms to effectively aggregate information from the local neighborhood of each atom. Crucially, E2Former operates on high-order tensors and employs tensor products to enrich the feature space while rigorously maintaining rotation and translation equivariances. We extended this architecture through the incorporation of our Long-Short-Range (LSR) message passing paradigm \cite{lilong}. The resulting E2Former-LSR maintains the efficiency of local message-passing for fine-grained, short-range interactions (covalent bonds and repulsion) while fundamentally addressing non-local correlations. As shown in Figure \ref{fig:main:1}.c, this is achieved by segmenting the entire molecular system into chemically meaningful fragments through some empirical methods, such as BRICS~\cite{brics:degen2008art}, and then employing an alternative distant attention block that aggregates features from local atomic neighbors (Short Interaction) and remote, fragment-level neighbors (Long Interaction), respectively. In this way, E2Former-LSR enables each atom to capture comprehensive, holistic molecular information, thereby overcoming the fixed-cutoff limitation while retaining high computational efficiency, a critical feature for scaling to macromolecular systems. Detailed information regarding the architecture and implementation of E2Former-LSR is provided in the Methods section and Supplementary Information.

\subsection*{Long-Range Interaction Dissociation Test}

\begin{figure}[htbp]
  \centering
  \includegraphics[width=1.0\linewidth]{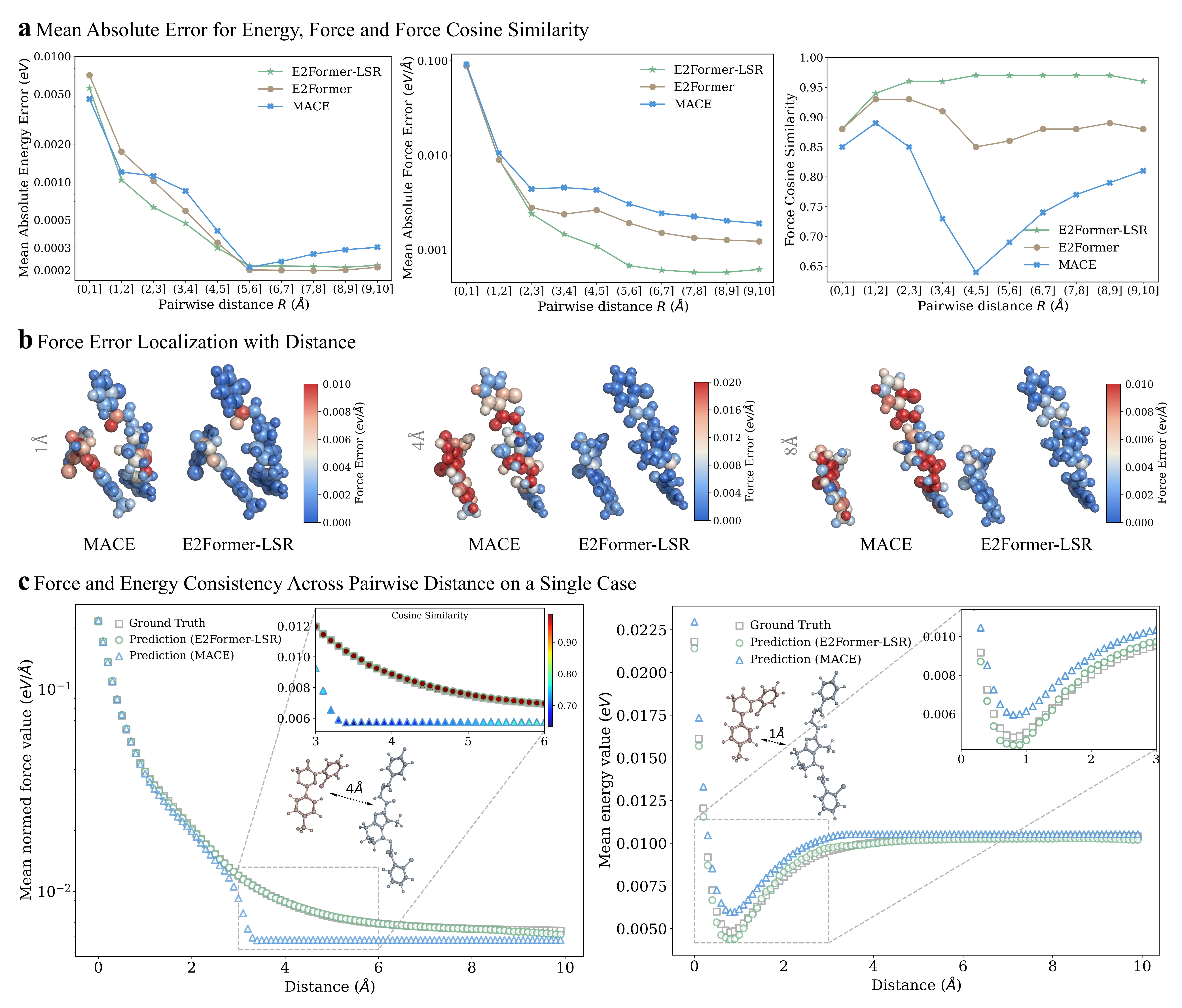}
  \caption{\textbf{Evaluation of long-range force and energy prediction accuracy on the Di-Molecule Dissociation dataset.} 
  \textbf{a}. Distance-resolved evaluation of MAEs for energy and force, together with CS$_f$. E2Former-LSR achieves smoother decay and higher directional consistency across all distances, especially in the transition region between short-range repulsion and long-range interaction. 
  \textbf{b}. Spatial visualization of force error magnitude for representative molecular dimers at three separation distances (1$\text{\r{A}}$, 4$\text{\r{A}}$, and 8$\text{\r{A}}$). Compared with MACE, E2Former-LSR maintains substantially lower and more uniformly distributed errors as molecular interaction decreases. 
  \textbf{c}. Single-system dissociation trajectory demonstrating prediction smoothness and physical continuity. The model outputs are evaluated across 100 frames with separation distances ranging from 0.2$\text{\r{A}}$ to 10$\text{\r{A}}$ in 0.1$\text{\r{A}}$ increments. E2Former-LSR accurately captures the asymptotic decay of forces and preserves a continuous and smooth potential energy surface, while MACE exhibits discontinuities and elevated errors at intermediate ranges.}
  \label{fig:results:2}
\end{figure}

We first utilized the \textit{Di-Molecule Dissociation Dataset} and the \textit{Long-Range Interaction Dissociation Test} to rigorously assess the capability of MLFFs to predict smooth potential energy surfaces and accurately capture non-covalent interactions over extended distances. This test involves systematically generating dissociation profiles for 100 molecules, stepping the intermolecular distance from $0.2\text{\r{A}}$ (near contact) to $10.1\text{\r{A}}$ in precise $0.1\text{\r{A}}$ increments.  The motivation of this test is twofold: (1) it explicitly verifies the model's ability to maintain a continuous and physically smooth Potential Energy Surface (PES) and corresponding forces as interactions transition from strong short-range repulsion to weak long-range non-covalent attraction; and (2) it provides a direct, fine-grained measure of how well the model handles the asymptotic decay of forces, which is fundamentally tied to long-range effects often missed by fixed-cutoff models.

We evaluated our proposed E2Former-LSR against leading local models, Allegro and the latest MACE-large \cite{mace-off:kovacs2025mace} (noted as MACE), DPA-2 \cite{dpa2:zhang2024dpa}, alongside the standard E2Former architecture (E2Former-Base). To better characterize the correlation of force directions, we additionally incorporated the cosine similarity of force (CS$_f$) metric in our analysis.

The results are shown as binned error curves in Figure \ref{fig:results:2}.a, which displays MAE for energy, force, and CS$_f$ segmented in $1\text{\r{A}}$ intervals. 
Analysis of the distance-binned errors reveals distinct performance regimes: in the short-range regime (e.g., the di-molecule distance $R \leq 2\text{\r{A}}$), where Pauli repulsion and covalent interactions dominate, local models exhibit prediction errors comparable to E2Former-LSR. Crucially, as the intermolecular distance increases (e.g., $R>2\text{\r{A}}$), the performance advantage of E2Former-LSR becomes significantly more pronounced, demonstrating substantial improvements in accuracy for both energy and forces.

The qualitative implications of force error are highlighted in Figure \ref{fig:results:2}.b, which visualizes the per-atom force error distribution at specific separation distances. While both MACE-large and E2Former-LSR maintain low error control at a distance of $1\text{\r{A}}$, the MACE-large error visibly increases at $4\text{\r{A}}$. In contrast, E2Former-LSR sustains very high accuracy even when the separation is extended to $8\text{\r{A}}$, confirming the model's efficacy in learning the essential long-range component that is fundamentally missed by fixed-cutoff architectures as their capacity to fully capture molecular information diminishes with separation.

To illustrate qualitative smoothness when the two molecules decouple from each other, we plot the energy and force profiles for a representative molecular pair from the test set in Figure \ref{fig:results:2}.c. The E2Former-LSR prediction learns a remarkably smooth, physically continuous curve that faithfully reproduces the DFT reference data across the entire range of separation. This qualitative fidelity is supported by consistently excellent Mean Absolute Error (MAE) and high Cosine Similarity values, demonstrating a robust and transferable understanding of long-range interaction decay. In stark contrast, the MACE prediction exhibits clear force cutoff artifacts, particularly beyond $R > 3\text{\r{A}}$, where the force profile displays unphysical discontinuities and sharp jumps, evidenced by a marked drop in Cosine Similarity in the extended range. Furthermore, MACE's prediction for the interaction energy shows poorer consistency with the DFT reference compared to E2Former-LSR as the separation distance $R$ increases. Additional examples verifying this smooth behavior across diverse molecular pairs are provided in the Supplementary Information.
More numerical results can be found in Supplementary Information Table \ref{tab:1:dimol}.

\subsection*{Medium-Scale Protein Conformation Fidelity Test}
\begin{figure}[htbp]
  \centering
  \includegraphics[width=1.0\linewidth]{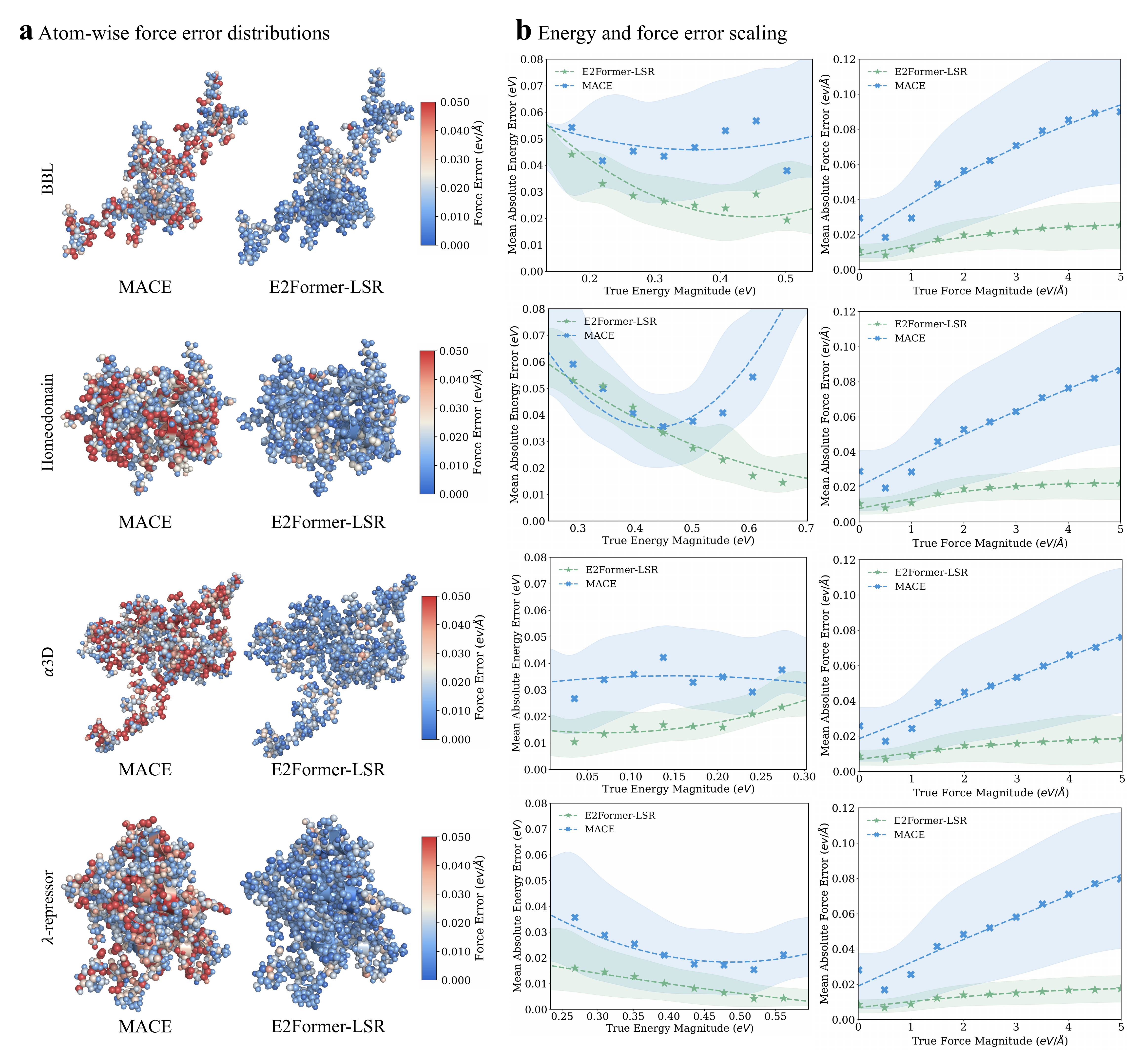}
  \caption{\textbf{Evaluation of model accuracy on large biomolecular systems in the Medium-Scale Protein Conformation Fidelity benchmark.} Four representative protein systems (BBL, Homeodomain, $\alpha$3D, and $\lambda$-repressor) extracted were used to assess robustness under realistic conformational complexity. 
  \textbf{a}. Atom-wise force error visualizations reveal that while MACE predictions exhibit spatially localized and context-dependent errors, E2Former-LSR produces substantially smoother and consistently lower-magnitude error distributions across all structures, reflecting its capacity for accurate non-local reasoning.
  \textbf{b}. Force-resolved analysis of mean absolute error as a function of true force magnitude. MACE demonstrates increased error variance sensitive to high-force intensity, whereas E2Former-LSR maintains uniformly low and stable error across the entire force spectrum. Corresponding energy error trends show that E2Former-LSR consistently preserves accuracy across high-dimensional conformational states, unlike MACE, which exhibits increased error sensitivity in higher energy regimes.} 
\label{fig:results:3}
\end{figure}

To rigorously evaluate model performance on realistic and complex biomolecular systems, we conducted the Medium-Scale Protein Conformation Fidelity (MS-PCF) Test. From D.E. Shaw’s extensive molecular dynamics (MD) trajectory dataset \cite{deshaw:lindorff2011fast}, we selected four large protein systems, ranging from 700 to 1200 atoms, after removing explicit solvent molecules to focus solely on the all-atom protein structures. This test is essential to validate the abilities of different architectures to maintain accuracy in highly complex, high-dimensional conformational spaces characteristic of native protein environments, where cooperative long-range non-covalent interactions are paramount.

We benchmarked E2Former-LSR against the leading models as in the previous test.
The numerical results can be found in Supplementary Information Table \ref{tab:2:deshaw}. Our model achieved an error reduction of up to 67\% for forces and up to 58\% for energy compared to MACE-large.
To gain a finer understanding of predictive fidelity beyond the overall MAE, we performed detailed atomic and conformational error analysis. The left of Figure \ref{fig:results:3} illustrates the per-atom force error on four representative proteins, comparing MACE-large and E2Former-LSR. MACE-large exhibits a pronounced increase in error for \emph{peripheral} atoms—those with fewer neighbors—underscoring its strong dependence on local connectivity. In sharp contrast, E2Former-LSR maintains excellent consistency, accurately predicting forces for both core and peripheral atoms.

Furthermore, to assess robustness in modeling unstable or highly strained conformations, we analyzed the error scaling against the magnitude of atomic forces and relative energy changes. As shown on the right of Figure \ref{fig:results:3}, the error of E2Former-LSR does not increase significantly with the amplitude of the forces or large relative energy changes. This stability under extreme conditions is crucial for accurately describing unstable systems and ensures that E2Former-LSR maintains a more reliable PES than local models, whose error scales more sharply with force magnitude.

\subsection*{Large-Molecule MD Trajectory Stability Test}

The ultimate validation for MLFFs lies in their performance in downstream molecular dynamics (MD) applications, which requires both accuracy and long-term stability of the potential energy surface (PES). To this end, we constructed a final benchmark derived from $10\text{ ps}$ MD trajectories spanning diverse chemical environments: pure water clusters, solvated inorganic salts ($\text{NaCl}$, $\text{NaOH}$, $\text{H}_2\text{SO}_4$ clusters), solvated organic molecules (Gln-Gly dipeptide and sucrose surrounded by water), and the complex $\text{ZIF-8}$ Metal-Organic Framework. For analysis, the first $70\%$ of each trajectory was used for model training, and the remaining $30\%$ was reserved for validation, test and dynamic analysis.

The quantitative results on the test set are presented in Table \ref{tab:3:md_traj}. Overall, E2Former-LSR demonstrates superior test metrics, particularly in the prediction of atomic forces. MACE, in contrast, exhibits competitive accurate performance in total energy prediction.

\begin{figure}[htbp]
  \centering
  \includegraphics[width=1.0\linewidth]{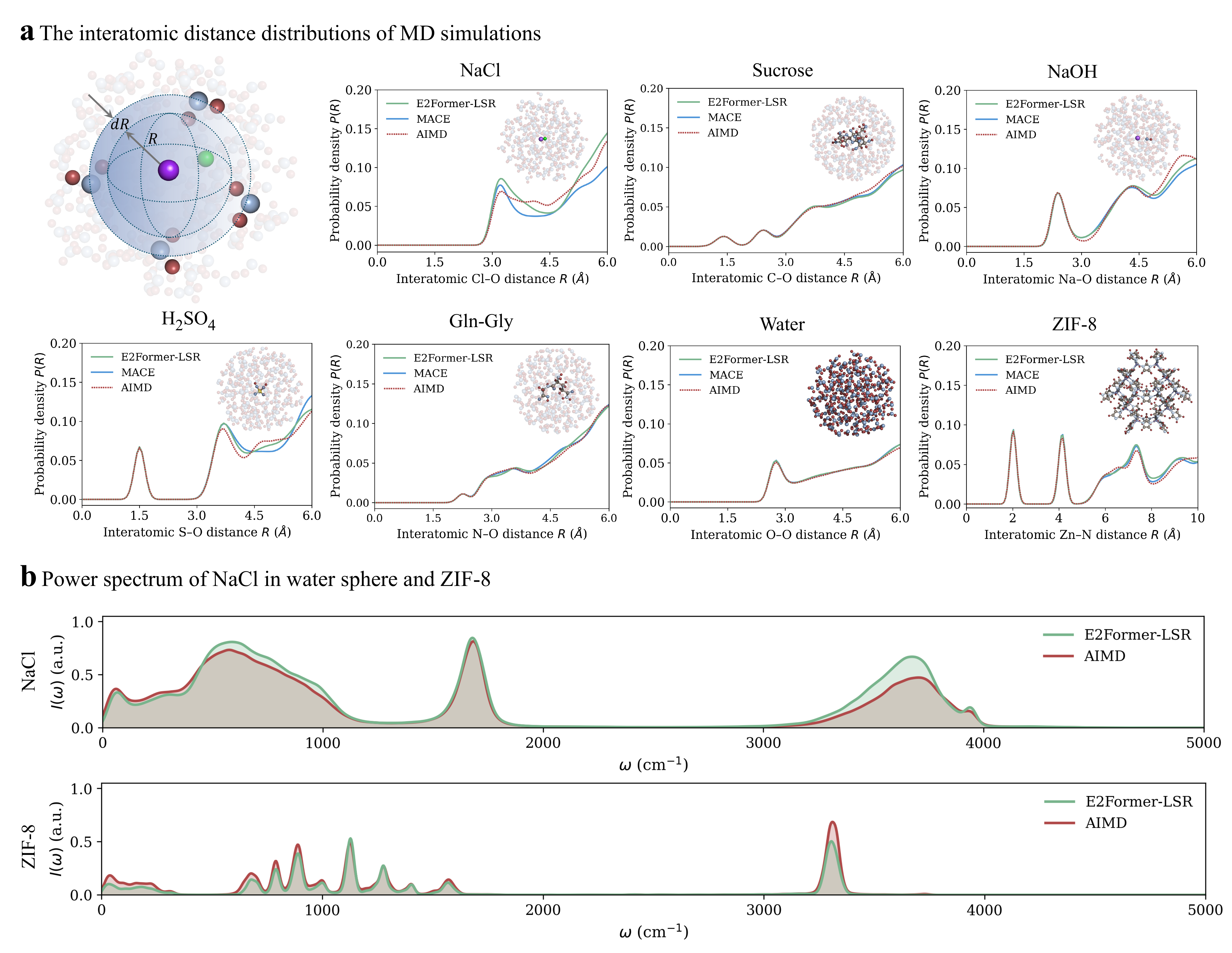}
\caption{\textbf{Validation of long-term dynamic stability and structural fidelity of E2Former-LSR and MACE.}
\textbf{a}. Interatomic atomic distance distribution derived from MD trajectories generated by E2Former-LSR and MACE, benchmarked against \textit{ab initio} MD (AIMD) references across seven representative large-molecule systems. These systems include pure water clusters, solvated inorganic salts ($\text{NaCl}$, $\text{NaOH}$, and $\text{H}_2\text{SO}_4$), solvated organic molecules (Gln-Gly and sucrose), and the $\text{ZIF-8}$ Metal-Organic Framework. The probability density presented are tailored to specific structural features (e.g., $\text{Cl-O}$, $\text{S-O}$, $\text{Zn-N}$) to reflect local ordering. Both MLFFs accurately capture the first coordination shell (first peak), but E2Former-LSR maintains a consistently tighter agreement with the AIMD reference, particularly across the medium-range correlations ($4\text{\r{A}}$ to $6\text{\r{A}}$). 
\textbf{b}. Power spectrum computed from the velocity autocorrelation functions for the solvated $\text{NaCl}$ cluster and the $\text{ZIF-8}$ framework. E2Former-LSR accurately reproduces the full vibrational spectrum, faithfully aligning the peak positions (representing collective modes) and spectral shapes with the AIMD reference. These results confirm that the superior force fidelity of E2Former-LSR directly translates to stable, high-accuracy structural and dynamic properties in extended MD simulations.}
  \label{fig:results:4}
\end{figure}

The advantage of high force accuracy becomes evident when evaluating the derived dynamic properties. We assessed the system-specific local structure by evaluating the interatomic distance distribution across the tested systems, displayed in Figure \ref{fig:results:4}.a. Specifically, we measured key inter-atomic distributions tailored to each system's composition, such as the \text{Cl-O} distribution in the \text{NaCl} cluster, the \text{C-O} distribution in the \text{Sucrose} cluster, the \text{Na-O} distribution in the \text{NaOH} cluster, the \text{S-O} distribution in the $\text{H}_2\text{SO}_4$ cluster, the \text{N-O} distribution in the \text{Gln-Gly} cluster, the {O-O} distribution in the \text{water} cluster and the  and the \text{Zn-N} distribution in the \text{ZIF-8} MOF. Both the E2Former-LSR and MACE models generally align well with the DFT reference, particularly in accurately reproducing the position and intensity of the first probability density peak, indicating reliable prediction of immediate local ordering. Critically, E2Former-LSR consistently demonstrates a tighter overall agreement with the DFT reference interatomic distance distribution. This confirms that its superior force fidelity directly translates to a more accurate representation of molecular structure and local ordering within dynamic simulations.

Furthermore, we examined the dynamical stability by comparing the power spectrum for the \text{NaCl} solution and \text{ZIF-8} systems, presented in Figure \ref{fig:results:4}.b. E2Former-LSR accurately aligns the vibrational peak positions with the DFT reference, and the amplitudes also show very close agreement, with only minor deviations. This demonstrates the model's capacity to maintain the correct high-dimensional dynamics over extended simulation times.

\begin{table}[h]
\caption{MAE for force (meV/\AA) and energies (meV) across seven systems.}
\label{tab:3:md_traj}
\centering
\setlength{\tabcolsep}{4pt}
\renewcommand{\arraystretch}{1.1}
\begin{tabularx}{\textwidth}{@{}l l *{4}{>{\centering\arraybackslash}X}@{}}
\toprule
\multirow{2}{*}{Molecule} & & \multicolumn{4}{c}{Models} \\
\cmidrule(lr){3-6}
 &  & E2Former-LSR & E2Former-Base & MACE-large & Allegro \\
\midrule

\multirow{2}{*}{\ce{H2SO4}}
  & Energy & \ColorCellMM{0.049}{0.205}{0.049}
           & \ColorCellMM{0.049}{0.205}{0.192}
           & \ColorCellMM{0.049}{0.205}{0.205}
           & \ColorCellMM{0.049}{0.205}{0.104} \\
  & Forces & \ColorCellMM{7.50}{24.65}{7.50}
           & \ColorCellMM{7.50}{24.65}{7.87}
           & \ColorCellMM{7.50}{24.65}{7.72}
           & \ColorCellMM{7.50}{24.65}{24.65} \\
\midrule

\multirow{2}{*}{\ce{NaCl}}
  & Energy & \ColorCellMM{0.062}{0.287}{0.062}
           & \ColorCellMM{0.062}{0.287}{0.101}
           & \ColorCellMM{0.062}{0.287}{0.287}
           & \ColorCellMM{0.062}{0.287}{0.110} \\
  & Forces & \ColorCellMM{6.46}{23.59}{6.64}
           & \ColorCellMM{6.46}{23.59}{7.46}
           & \ColorCellMM{6.46}{23.59}{6.98}
           & \ColorCellMM{6.46}{23.59}{23.59} \\
\midrule

\multirow{2}{*}{\ce{NaOH}}
  & Energy & \ColorCellMM{0.055}{0.168}{0.055}
           & \ColorCellMM{0.055}{0.168}{0.084}
           & \ColorCellMM{0.055}{0.168}{0.084}
           & \ColorCellMM{0.055}{0.168}{0.168} \\
  & Forces & \ColorCellMM{6.64}{21.68}{6.64}
           & \ColorCellMM{6.64}{21.68}{7.16}
           & \ColorCellMM{6.64}{21.68}{6.46}
           & \ColorCellMM{6.64}{21.68}{21.68} \\
\midrule

\multirow{2}{*}{Gln-Gly}
  & Energy & \ColorCellMM{0.062}{0.156}{0.084}
           & \ColorCellMM{0.062}{0.156}{0.144}
           & \ColorCellMM{0.062}{0.156}{0.062}
           & \ColorCellMM{0.062}{0.156}{0.156} \\
  & Forces & \ColorCellMM{8.15}{26.01}{8.48}
           & \ColorCellMM{8.15}{26.01}{9.11}
           & \ColorCellMM{8.15}{26.01}{8.15}
           & \ColorCellMM{8.15}{26.01}{26.01} \\
\midrule

\multirow{2}{*}{Sucrose}
  & Energy & \ColorCellMM{0.048}{0.147}{0.108}
           & \ColorCellMM{0.048}{0.147}{0.118}
           & \ColorCellMM{0.048}{0.147}{0.048}
           & \ColorCellMM{0.048}{0.147}{0.147} \\
  & Forces & \ColorCellMM{6.46}{22.98}{6.46}
           & \ColorCellMM{6.46}{22.98}{6.85}
           & \ColorCellMM{6.46}{22.98}{7.03}
           & \ColorCellMM{6.46}{22.98}{22.98} \\
\midrule

\multirow{2}{*}{\ce{Water}}
  & Energy & \ColorCellMM{0.091}{0.120}{0.091}
           & \ColorCellMM{0.091}{0.120}{0.113}
           & \ColorCellMM{0.091}{0.120}{0.120}
           & \ColorCellMM{0.091}{0.120}{0.110} \\
  & Forces & \ColorCellMM{6.33}{23.37}{6.33}
           & \ColorCellMM{6.33}{23.37}{7.03}
           & \ColorCellMM{6.33}{23.37}{6.77}
           & \ColorCellMM{6.33}{23.37}{23.37} \\
\midrule

\multirow{2}{*}{ZIF-8}
  & Energy & \ColorCellMM{0.075}{0.238}{0.075}
           & \ColorCellMM{0.075}{0.238}{0.182}
           & \ColorCellMM{0.075}{0.238}{0.238}
           & \ColorCellMM{0.075}{0.238}{0.118} \\
  & Forces & \ColorCellMM{4.73}{15.61}{4.73}
           & \ColorCellMM{4.73}{15.61}{5.51}
           & \ColorCellMM{4.73}{15.61}{6.68}
           & \ColorCellMM{4.73}{15.61}{15.61} \\
\bottomrule
\end{tabularx}
\end{table}

\section*{Discussions and Outlooks}

The reliance of contemporary machine learning force fields (MLFFs) on short-range locality has imposed a critical, system-size-dependent bottleneck on their applicability to macromolecular simulation. Our rigorous assessment of leading local models confirmed that force prediction error monotonically increases with system size (Figure \ref{fig:main:1}.b), revealing a fundamental architectural limitation inherent to fixed-cutoff approaches. This limitation, compounded by the scarcity of suitable benchmark data, has prevented the widespread adoption of MLFFs in realistic biophysical and materials modeling.

We resolved this challenge by introducing E2Former-LSR, an equivariant transformer architecture that explicitly integrates a Long-Short-Range message passing framework. This design successfully models interactions fundamentally inaccessible to traditional methods. The empirical results on our comprehensive \emph{MolLR25} benchmark suite demonstrate that this non-local paradigm successfully addresses the scaling problem: E2Former-LSR exhibits stable error scaling even for systems exceeding 1200 atoms (Figure \ref{fig:main:1}.b), while achieving up to an 30\% speedup compared to purely local models. Furthermore, its performance on dedicated long-range tests validates its enhanced physical fidelity; it robustly captures the smooth, physical decay of non-covalent potentials (Figure \ref{fig:results:2}.c) and maintains superior accuracy on complex, medium-scale protein conformations (Table \ref{tab:2:deshaw}), validating its ability to generalize to realistic, high-density environments.

The successful introduction of E2Former-LSR demonstrates that the primary obstacle to accurate macromolecular simulation by MLFFs is architectural, not merely data-related. This work paves the way for future MLFF development to prioritize the explicit and efficient treatment of non-local effects, enabling high-fidelity molecular dynamics simulations across entire biological and material domains that were previously only accessible via computationally prohibitive methods. This advance opens up new avenues for exploring complex conformational dynamics and large-scale assembly mechanisms.
\section*{Online Methods}

\subsection*{MolLR25 Data Preparation}

All reference DFT calculations were performed using the $\omega\text{B97X-D3}$/$\text{def2-SVP}$ level of theory \cite{weigend2005balanced,grimme2010consistent} to ensure accurate treatment of long-range van der Waals interactions. Calculations utilized a GPU-accelerated version of PySCF (GPU4PySCF) \cite{pyscf:sun2020recent,pyscf-gpu:pu2025enhancing}. For generating the MD Trajectory Dataset, \emph{ab initio} molecular dynamics (AIMD) was performed in the $\text{NVT}$ ensemble at $300\text{ K}$ using the Atomic Simulation Environment (ASE) \cite{ase:larsen2017atomic} with GPU4PySCF as the calculator.

Our dataset is a systematically designed, high-fidelity benchmark suite comprising three categories, tailored to stress-test long-range interaction learning across diverse molecular domains and length scales:

\begin{itemize}
\item \textbf{Di-Molecule Dissociation Dataset.} We constructed 4950 molecular dimers from 100 organic molecules sourced from PubChemQC \cite{pubchem:nakata2023pubchemlogc}. DFT calculations were performed by systematically increasing the inter-monomer separation from $0.1\text{\r{A}}$ to $10.1\text{\r{A}}$ ($0.1\text{\r{A}}$ increments). This high-resolution setup (totaling $\approx 500,000$ DFT data points) provides a controlled environment to verify the smooth, asymptotic decay of non-bonded forces.

\item \textbf{Medium-Scale Protein Dataset.} Targeting realistic biophysical scenarios, we curated static protein snapshots ($700$ to $1200$ atoms) from D. E. Shaw Research MD trajectories \cite{deshaw:lindorff2011fast}. Extracted conformations were re-evaluated via DFT, yielding over 48,000 high-quality energy and force labels. This dataset spans interaction distances up to $70\text{\r{A}}$, crucial for probing non-local effects in structured biomolecules.

\item \textbf{MD Trajectory Dataset.} To assess dynamic stability, we constructed $10\text{ ps}$ AIMD trajectories for diverse large systems (exceeding $500$ atoms), including water clusters, solvated small molecules, and a $\text{ZIF-8}$ MOF. This component examines long-range consistency, energy conservation, and trajectory stability across extended simulations.
\end{itemize}

The rich composition and statistical properties of the $\emph{MolLR25}$ suite are detailed in Figure \ref{fig:main:1}.a and Supplementary Information. Together, these high-accuracy datasets form a purpose-built foundation for evaluating long-range-aware MLFFs.

\subsection*{E2Former-LSR: A Unified Long–Short Range Equivariant Framework}
\label{sec:lsr-e2former}

We developed \textbf{E2former-LSR}, a unified $\mathrm{SO}(3)$-equivariant neural architecture that integrates \emph{Long–Short Range Message Passing} (LSR-MP) with an \emph{E2Former} backbone to capture both local and nonlocal interactions in molecular systems. 
Given an $n$-atom system with atomic numbers $Z\in\mathbb{N}^{n}$ and Cartesian coordinates $\mathbf{P}\in\mathbb{R}^{n\times3}$, E2Former-LSR constructs three complementary representations:
(i) a \emph{short-range atomic graph} $G_{\text{short}}$ with cutoff $r_{\text{short}}$ for dense many-body interactions,
(ii) a \emph{fragment set} $U$ capturing chemically coherent substructures, and
(iii) a \emph{long-range atom–fragment bipartite graph} $G_{\text{long}}$ with $r_{\text{long}}\gg r_{\text{short}}$ to model nonlocal couplings. All message-passing operations are implemented by \textbf{E2Former layers} employing \emph{Wigner-$6j$–based equivariant attention} and \emph{node-wise Wigner convolution}, which preserve strict $\mathrm{SO}(3)$ equivariance \cite{murnaghan1938analysis} while achieving linear complexity with respect to graph sparsity. 
Final short- and long-range representations are combined through a late-fusion \cite{baltruvsaitis2018multimodal} step before property prediction.

\subsubsection*{Symbols and Notation}
\label{subsec:symbols}
\begin{tabular}{ll}
\toprule
Symbol & Meaning \\
\midrule
$Z_i$ & Atomic number of atom $i$ (type-0 input).\\
$\mathbf{p}_i\in\mathbb{R}^3$ & Cartesian position of atom $i$; $\mathbf{P}_j$ is fragment center.\\
$G_{\text{short}}=(V,E_{\text{short}})$ & Short-range radius graph with cutoff $r_{\text{short}}$.\\
$U$, $S(j)$ & Fragment index set, and atom set of fragment $j$.\\
$G_{\text{long}}=(V,U,E_{\text{long}})$ & Atom--fragment bipartite graph with cutoff $r_{\text{long}}$.\\
$\mathcal{L}=\{0,\dots,L_{\max}\}$ & Angular orders of SO(3) irreps.\\
$\mathbf{h}^{(t)}_{i,\ell}$ & Order-$\ell$ irrep features of atom $i$ at layer $t$ (short-range).\\
$\mathbf{H}^{(t)}_{j,\ell}$ & Order-$\ell$ irrep features of fragment $j$ at layer $t$.\\
$\mathbf{x}^{(t)}_{i,\ell}$, $\boldsymbol{\mu}^{(t)}_{j,\ell}$ & Long-range atom/fragment irrep states at layer $t$.\\
$\mathbf{Y}_\ell(\hat{\mathbf{r}})$ & Real spherical harmonics of direction $\hat{\mathbf{r}}$.\\
$\otimes$, $\langle\cdot\rangle_{(0)}$ & CG tensor product, and projection to scalar irrep.\\
$\{6j\}$ & Wigner $6j$ symbol used for recoupling CG paths.\\
$L_{\text{short}},L_{\text{long}}$ & Short/long-range layer counts.\\
$H$ & \#attention heads per order.\\
$d_\ell$ & Channel multiplicity for order-$\ell$ irrep block.\\
\bottomrule
\end{tabular}

\subsubsection*{Fragmentation Module}
\label{subsec:frag}

To capture mesoscale coherence and chemical context, E2Former-LSR introduces a \textbf{fragmentation module} that partitions atoms into fragments $\{1,\dots,|U|\}$ using either chemically informed decomposition (e.g., BRICS~\cite{brics:degen2008art, liu2017break}) or geometry-based clustering (e.g., k-means or k-nearest neighbors algorithms~\cite{gnanadesikan2011methods, cover1967nearest}). 
Each fragment $u$, defined by its associated atom set $S(u)\subseteq V$, is represented by an $\mathrm{SO}(3)$-invariant geometric center:
\begin{equation}
\mathbf{P}_u = \sum_{i\in S(u)} \gamma_i\,\mathbf{p}_i, 
\qquad 
\sum_{i\in S(u)}\gamma_i=1,
\label{eq:frag_center}
\end{equation}
where $\gamma_i$ denotes the normalized weighting coefficient defining the contribution of atom $i$ to the fragment center. 
In practice, $\gamma_i$ can be derived from the clustering procedure: for k-means, the cluster centroid itself serves as $\mathbf{P}_u$, whereas for BRICS decomposition, $\mathbf{P}_u$ corresponds to the average position of all atoms belonging to the same chemically defined subunit.
This construction provides a smooth and rotation-invariant mapping from atomic coordinates to fragment-level representations, facilitating stable coupling between local and long-range modules.

In this paper, chemically aware fragmentation, i.e., BRICS, preserves bonding patterns and reduces ``bond-cut'' artifacts compared to purely geometric clustering, leading to fragment features that serve as physically meaningful carriers of long-range information.

\subsubsection*{Short-Range Module}

The Short-Range Module captures local many-body and angular interactions. 
Conceptually, the short-range block learns local potential energy surfaces by propagating orientation-aware messages within each atom’s neighborhood, thereby encoding many-body correlations at quantum accuracy. 
Each atom node is initialized with irreducible representation (irrep\cite{wigner2012group}) features derived from atomic embeddings, and then, a stack of $L_{\text{short}}$ \textbf{E2Former layers} updates these features as shown in Eq. \eqref{eq:init_ir} and Eq. \eqref{eq:short_e2former}. The E2Former design shifts Clebsch--Gordan (CG) tensor products\cite{edmonds1996angular} from edges to nodes via Wigner-$6j$\cite{wigner2012group} recoupling, yielding linear complexity while maintaining high-order geometric expressivity.

\begin{equation}
\mathbf{h}^{(0)}_{i,0}=\mathrm{Embed}(Z_i), 
\qquad
\mathbf{h}^{(0)}_{i,\ell}=\mathbf{0}\;(\ell\ge1).
\label{eq:init_ir}
\end{equation}

\begin{equation}
\mathbf{h}^{(t+1)}_{i,\ell} 
=\mathrm{E2Layer}_{\text{short}}\!\Big(
\mathbf{h}^{(t)}_{i,\cdot},\;
\{\mathbf{h}^{(t)}_{j,\cdot},\mathbf{Y}_\cdot(\hat{\mathbf{r}}_{ij})\}_{j\in\mathcal{N}_{\text{short}}(i)}
\Big),
\quad
t=0,\dots,L_{\text{short}}{-}1,
\label{eq:short_e2former}
\end{equation}
where $\mathbf{Y}_\ell(\hat{\mathbf{r}}_{ij})$ are real spherical harmonics\cite{sphericalHamonics}, and ${\mathcal{N}_{\text{short}}(i)}:=\big\{(i,j)\,:\,\|\mathbf{p}_i-\mathbf{p}_j\|\le r_{\text{short}},j\in V\big\}$ construct a local neighborhood within a radius graph.

\subsubsection*{Long-Range Module}
\label{subsec:long}

Long-range interactions—such as polarization\cite{polarization}, electrostatic coupling\cite{polarization}, and through-space correlation—are modeled through a \textbf{bipartite atom–fragment graph} defined as
\begin{equation}
\mathcal{N}_{\text{long}}
=\big\{(i,u)\in V\times U\,:\,\|\mathbf{p}_i-\mathbf{P}_u\|\le r_{\text{long}}\big\},
\qquad 
r_{\text{long}}\gg r_{\text{short}}.
\label{eq:elong}
\end{equation}
This construction establishes directional connections between atomic nodes $V$ and fragment nodes $U$, enabling efficient information exchange across distant regions of the system without forming a fully connected atomic graph.

The long-range module is initialized using atomic representations propagated from the short-range block and fragment-level irreducible-representation (irrep) descriptors computed from the fragmentation stage through mean pooling:
\begin{equation}
\mathbf{x}^{(0)}_{i,\ell}=\mathbf{h}^{(L_{\text{short}})}_{i,\ell}, 
\qquad 
\boldsymbol{\mu}^{(0)}_{u,\ell} = \mathrm{meanpool}_{i\in S(u)} \mathbf{h}_{i,\ell}^{(L_{short})},
\label{eq:long_init}
\end{equation}
Subsequent layers perform \textbf{bipartite E2Attention}, where each atom attends to fragment nodes via the spherical harmonics of the relative orientation 
$\hat{\mathbf{r}}_{iu}=(\mathbf{P}_u-\mathbf{p}_i)/\|\mathbf{P}_u-\mathbf{p}_i\|$. 
Analogous to the local update rule in Eq.~\ref{eq:short_e2former}, the long-range propagation follows
\begin{equation}
\mathbf{x}^{(t+1)}_{i,\ell} 
= 
\mathrm{E2Layer}_{\text{long}}\!\Big(
\mathbf{x}^{(t)}_{i,\cdot},\;
\{\boldsymbol{\mu}^{(t)}_{u,\cdot},\mathbf{Y}_\cdot(\hat{\mathbf{r}}_{iu})\}_{u\in\mathcal{N}_{\text{long}}(i)}
\Big),
\quad
t=0,\dots,L_{\text{long}}{-}1,
\label{eq:long_stack}
\end{equation}
where $\mathcal{N}_{\text{long}}(i)$ denotes the set of fragment neighbors of atom $i$. 
This formulation mirrors the short-range update scheme while extending the receptive field to nonlocal atom–fragment interactions in an $\mathrm{SO}(3)$-equivariant manner, preserving both geometric consistency and computational linearity with respect to system size.

The resulting long-range atomic features $\mathbf{x}^{(L_{\text{long}})}_i$ are combined with the short-range embeddings through a \textbf{late-fusion} operation:
\begin{equation}
\mathbf{z}_i 
= 
\mathrm{Fuse}\!\left(
\mathbf{h}^{(L_{\text{short}})}_i,\,
\mathbf{x}^{(L_{\text{long}})}_i
\right),
\label{eq:late_fusion}
\end{equation}
producing unified and multi-scale representations that simultaneously encode local atomic physics, fragment-level chemical context, and nonlocal field effects for downstream property prediction.

\subsubsection*{Property Heads and Training Objective}
\label{subsec:heads}

From the fused atomic representations $\mathbf{z}_i$, E2Former-LSR predicts both the total molecular energy and per-atom forces in a physically consistent manner:
\begin{equation}
\widehat{E}
=\sum_{i\in V} g(\mathbf{z}_{i,0}),
\qquad
\widehat{\mathbf{F}}_i 
= -\frac{\partial \widehat{E}}{\partial \mathbf{p}_i},
\label{eq:energy_forces}
\end{equation}
where $g$ is a scalar head operating on the invariant ($\ell=0$) components of the final atomic features. 
By deriving forces as analytical gradients of the predicted energy, the model preserves exact energy–force consistency and differentiability with respect to atomic coordinates.

The network is optimized end-to-end using a joint energy–force objective:
\begin{equation}
\mathcal{L}
=\lambda_E\|\widehat{E}-E\|_1
+\lambda_F\,\frac{1}{n}\sum_i
\|\widehat{\mathbf{F}}_i-\mathbf{F}_i\|_1,
\label{eq:loss}
\end{equation}
where $\lambda_E$ and $\lambda_F$ control the relative weighting between energy and force terms. 
This formulation enforces accurate energy prediction while ensuring that the learned potential yields physically faithful force fields through automatic differentiation.

\subsubsection*{Experimental Settings}

\paragraph{E2Former-LSR Hyper-Parameters}
\label{subsec:exp}

\begin{tabularx}{\textwidth}{l l X}
\toprule
Name & Description & Typical value(s) \\
\midrule
$r_{\text{short}}$ & Short-range cutoff & $5\,\mathrm{\AA}$ \\
$r_{\text{long}}$ & Long-range cutoff (bipartite) & $\approx 15\,\mathrm{\AA}$ \\
$L_{\max}$ & Max angular order & $1$ or $2$ \\
$H$ & Attention heads per order & $4\text{--}8$ \\
$L_{\text{short}}$ & \#E2Former layers on $G_{\text{short}}$ & $4$ \\
$L_{\text{long}}$ & \#E2Former layers on $G_{\text{long}}$ & $2$ \\
$\{d_\ell\}$ & Channels per irrep order & $d_0{=}256,\ d_1{=}128,\ d_2{=}96$ \\
Optimizer & AdamW $(\beta_1,\beta_2)$, weight decay & $(0.9,0.999)$,\ $10^{-4}$ \\
LR schedule & Base LR, cosine decay, warmup & $1\!\times\!10^{-4}$,\ $5\%$ warmup \\
Loss weights & $(\lambda_E,\lambda_F)$ in Eq.~\eqref{eq:energy_forces} & $(1,\,100)$ \\
\bottomrule
\end{tabularx}

\paragraph{Baseline Models}
To ensure fair comparison across different machine learning force-field architectures, we adopt consistent training and preprocessing pipelines for all baseline models considered in this work, including Allegro, MACE-Large, and DPA-2. Each model is configured following its official reference implementation, with only minimal adjustments to cutoff radius and feature widths to align with the molecular systems evaluated in our experiments. All baselines operate on local neighborhoods constructed with short-range cutoff of $5$\,\AA.

The equivariant baselines (Allegro and MACE-Large) follow their established implementations and employ spherical-harmonic features up to second order, with standard hidden-channel sizes and the default per-atom readout MLP used in their original designs. MACE-Large is instantiated with its higher-order correlation design, while Allegro incorporates its optimized two-body radial embedding and tensor-product interaction stack. To complement these equivariant models, we additionally include DPA-2, which combines a long-range attention branch with a short-range SE-based equivariant branch, unified through a shared fitting network. All baseline models are trained using the same optimizer, identical force-dominant loss weighting, and a consistent batching and neighbor-list construction strategy. Detailed configurations of all baseline models (Allegro, MACE-Large and DPA-2) are provided in Supplementary Information under \emph{Baseline Models Configuration}.

\subsubsection*{Summary}
\label{subsec:summary}

E2Former-LSR provides a unified $\mathrm{SO}(3)$-equivariant framework that seamlessly integrates local, fragment-level, and long-range interactions within molecular systems. 
By combining \emph{Wigner-$6j$–based equivariant attention} for efficient tensor recoupling, \emph{fragment-aware coarse-graining} for chemically interpretable representations, and \emph{bipartite long-range message passing} for scalable nonlocal modeling, 
the framework captures multi-scale physical correlations with quantum-level accuracy while maintaining near-linear computational scaling. 
This design bridges fine-grained atomic physics with coarse-grained chemical context, enabling transferable and data-efficient force-field learning across both molecular and condensed-phase systems.

\backmatter

\section*{Data and Code Availability}
The complete MolLR25 dataset, including all corresponding test code and trained model parameters, can be found on GitHub and Hugging Face to promote open science and reproducibility.
\begin{itemize}
    \item Code: https://github.com/IQuestLab/UBio-MolFM
    \item Data: https://huggingface.co/datasets/IQuestLab/UBio-MolLR25
    \item Model: https://huggingface.co/IQuestLab/UBio-E2Former-LSR
\end{itemize}

\section*{Acknowledgements}
We thank Dr. Han Yang for the helpful discussions. XW was supported by the Zhongguancun Academy under the Internal Research Grant No. C20250501. 

\bibliography{sn-bibliography}

@article{uma:wood2025family,
  title={{UMA}: A Family of Universal Models for Atoms},
  author={Wood, Brandon M and Dzamba, Misko and Fu, Xiang and Gao, Meng and Shuaibi, Muhammed and Barroso-Luque, Luis and Abdelmaqsoud, Kareem and Gharakhanyan, Vahe and Kitchin, John R and Levine, Daniel S and others},
  journal={arXiv preprint arXiv:2506.23971},
  year={2025}
}

@article{omol25:levine2025open,
  title={The open molecules 2025 (omol25) dataset, evaluations, and models},
  author={Levine, Daniel S and Shuaibi, Muhammed and Spotte-Smith, Evan Walter Clark and Taylor, Michael G and Hasyim, Muhammad R and Michel, Kyle and Batatia, Ilyes and Cs{\'a}nyi, G{\'a}bor and Dzamba, Misko and Eastman, Peter and others},
  journal={arXiv preprint arXiv:2505.08762},
  year={2025}
}

@article{deepmd:wang2018deepmd,
  title={{DeePMD-kit}: A deep learning package for many-body potential energy representation and molecular dynamics},
  author={Wang, Han and Zhang, Linfeng and Han, Jiequn and others},
  journal={Comput. Phys. Commun.},
  volume={228},
  pages={178--184},
  year={2018},
  publisher={Elsevier}
}

@article{deepmd2:zeng2023deepmd,
  title={{DeePMD-kit v2}: A software package for deep potential models},
  author={Zeng, Jinzhe and Zhang, Duo and Lu, Denghui and Mo, Pinghui and Li, Zeyu and Chen, Yixiao and Rynik, Mari{\'a}n and Huang, Li’ang and Li, Ziyao and Shi, Shaochen and others},
  journal={J. Chem. Phys.},
  volume={159},
  number={5},
  year={2023},
  publisher={AIP Publishing}
}

@article{deepmd3:zeng2025deepmd,
  title={{DeePMD-kit v3}: A multiple-backend framework for machine learning potentials},
  author={Zeng, Jinzhe and Zhang, Duo and Peng, Anyang and Zhang, Xiangyu and He, Sensen and Wang, Yan and Liu, Xinzijian and Bi, Hangrui and Li, Yifan and Cai, Chun and others},
  journal={J. Chem. Theory Comput.},
  volume={21},
  number={9},
  pages={4375--4385},
  year={2025},
  publisher={ACS Publications}
}

@article{uni-mol:zhouuni,
  title={{Uni-Mol}: A Universal {3D} Molecular Representation Learning Framework},
  author={Zhou, Gengmo and Gao, Zhifeng and Ding, Qiankun and Zheng, Hang and Xu, Hongteng and Wei, Zhewei and Zhang, Linfeng and Ke, Guolin},
  journal={The Eleventh International Conference on Learning Representations},
  year={2023},
}

@article{uni-mol2:ji2024uni,
author = {Ji, Xiaohong and Wang, Zhen and Gao, Zhifeng and Zheng, Hang and Zhang, Linfeng and Ke, Guolin and E, Weinan},
title = {Exploring molecular pretraining model at scale},
year = {2024},
isbn = {9798331314385},
publisher = {Curran Associates Inc.},
address = {Red Hook, NY, USA},
journal = {Proceedings of the 38th International Conference on Neural Information Processing Systems},
articleno = {1489},
numpages = {23},
location = {Vancouver, BC, Canada},
series = {NIPS '24},
}

@article{dpa:zhang2024pretraining,
  title={Pretraining of attention-based deep learning potential model for molecular simulation},
  author={Zhang, Duo and Bi, Hangrui and Dai, Fu-Zhi and Jiang, Wanrun and Liu, Xinzijian and Zhang, Linfeng and Wang, Han},
  journal={npj Comput. Mater.},
  volume={10},
  number={1},
  pages={94},
  year={2024},
  publisher={Nature Publishing Group UK London}
}

@article{dpa2:zhang2024dpa,
  title={{DPA-2}: A large atomic model as a multi-task learner},
  author={Zhang, Duo and Liu, Xinzijian and Zhang, Xiangyu and Zhang, Chengqian and Cai, Chun and Bi, Hangrui and Du, Yiming and Qin, Xuejian and Peng, Anyang and Huang, Jiameng and others},
  journal={npj Comput. Mater.},
  volume={10},
  number={1},
  pages={293},
  year={2024},
  publisher={Nature Publishing Group UK London}
}

@article{mace-off:kovacs2025mace,
  title={{MACE-OFF}: Short-range transferable machine learning force fields for organic molecules},
  author={Kov{\'a}cs, D{\'a}vid P{\'e}ter and Moore, J Harry and Browning, Nicholas J and Batatia, Ilyes and Horton, Joshua T and Pu, Yixuan and Kapil, Venkat and Witt, William C and Magdau, Ioan-Bogdan and Cole, Daniel J and others},
  journal={J. Am. Chem. Soc.},
  volume={147},
  number={21},
  pages={17598--17611},
  year={2025},
  publisher={ACS Publications}
}

@article{pubchem:nakata2023pubchemlogc,
  title={{PubChemQC B3LYP/6-31G*//PM6} data set: The electronic structures of 86 million molecules using {B3LYP/6-31G*} calculations},
  author={Nakata, Maho and Maeda, Toshiyuki},
  journal={J. Chem. Inf. Model.},
  volume={63},
  number={18},
  pages={5734--5754},
  year={2023},
  publisher={ACS Publications},
  doi={10.1021/acs.jcim.3c00424}
}

@article{pyscf-gpu:pu2025enhancing,
  title={Enhancing {PySCF}-based quantum chemistry simulations with modern hardware, algorithms, and {Python} tools},
  author={Pu, Zhichen and Sun, Qiming},
  journal={APL Comput. Phys.},
  volume={1},
  number={1},
  pages={016101},
  year={2025},
  publisher={AIP Publishing LLC}
}

@article{pyscf:sun2020recent,
  title={Recent developments in the {PySCF} program package},
  author={Sun, Qiming and Zhang, Xing and Banerjee, Sandeep and Bao, Peng and Barbry, Marcel and Blunt, Nathan S and Bogdanov, Nikita A and Booth, George H and Chen, Jia and Cui, Zhong Hao and Eriksen, Jens J},
  journal={J. Chem. Phys.},
  volume={153},
  number={2},
  pages={024109},
  year={2020},
  publisher={AIP Publishing},
  doi={10.1063/5.0019206}
}

@article{ase:larsen2017atomic,
  title={The Atomic Simulation Environment--A {Python} library for working with atoms},
  author={Larsen, Ask Hjorth and Mortensen, Jens J{\o}rgen and Blomqvist, Jakob and Castelli, Ivano E. and Christensen, Kasper S. and Du{\l}ak, Marcin and Friis, Jakob and Groenbeck, Morten and Jonson, Brian and Knees, Thomas and Lehtovaara, Lauri and Mortensen, Steen S and Olsen, Asger and Schi{\o}tz, Jakob and Strange, Mikael and Taylor, Kathleen and Tran, Steen and Walter, I-Ping},
  journal={J. Phys. Condens. Matter},
  volume={29},
  number={27},
  pages={273002},
  year={2017},
  publisher={IOP Publishing},
  doi={10.1088/1361-648X/aa680e}
}

@article{brics:degen2008art,
  title={On the art of compiling and using 'drug-like' chemical fragment spaces},
  author={Degen, Jorg and Wegscheid-Gerlach, Christof and Zaliani, Andrea and Rarey, Matthias},
  journal={ChemMedChem},
  volume={3},
  number={10},
  pages={1503},
  year={2008}
}

@article{grimme2010consistent,
  title={A consistent and accurate ab initio parametrization of density functional dispersion correction ({DFT-D}) for the 94 elements {H-Pu}},
  author={Grimme, Stefan and Antony, Jens and Ehrlich, Stephan and Krieg, Helge},
  journal={J. Chem. Phys.},
  volume={132},
  number={15},
  year={2010},
  publisher={AIP Publishing}
}

@article{lilong,
  title={Long-Short-Range Message-Passing: A Physics-Informed Framework to Capture Non-Local Interaction for Scalable Molecular Dynamics Simulation},
  author={Li, Yunyang and Wang, Yusong and Huang, Lin and Yang, Han and Wei, Xinran and Zhang, Jia and Wang, Tong and Wang, Zun and Shao, Bin and Liu, Tie-Yan},
  booktitle={The Twelfth International Conference on Learning Representations},
  year={2024}
}

@article{deshaw:lindorff2011fast,
  title={How fast-folding proteins fold},
  author={Lindorff-Larsen, Kresten and Piana, Stefano and Dror, Ron O and Shaw, David E},
  journal={Science},
  volume={334},
  number={6055},
  pages={517--520},
  year={2011},
  publisher={American Association for the Advancement of Science}
}

@article{weigend2005balanced,
  title={Balanced basis sets of split valence, triple zeta valence and quadruple zeta valence quality for {H} to {Rn}: Design and assessment of accuracy},
  author={Weigend, Florian and Ahlrichs, Reinhart},
  journal={Phys. Chem. Chem. Phys. },
  volume={7},
  number={18},
  pages={3297--3305},
  year={2005},
  publisher={Royal Society of Chemistry}
}

@article{liaoequiformerv2,
  title={EquiformerV2: Improved Equivariant Transformer for Scaling to Higher-Degree Representations},
  author={Liao, Yi-Lun and Wood, Brandon M and Das, Abhishek and Smidt, Tess},
  booktitle={The Twelfth International Conference on Learning Representations},
  year={2024},
editor={},
month={},
}

@article{schnet,
  author = {Sch{\"u}tt, Kristof T and et al.},
  title = {{SchNet}: A continuous-filter convolutional neural network for modeling quantum interactions},
  journal = {Advances in Neural Information Processing Systems},
  year = {2017}
}

@article{sgdml,
  author = {Chmiela, Stefan and et al.},
  title = {Machine learning of accurate energy-conserving molecular force fields},
  journal = {Sci. Adv.},
  year = {2017}
}

@article{equiformer,
  author = {Bao, Zhen and et al.},
  title = {Equiformer: Equivariant Graph Attention Transformer for {3D} Atomistic Representations},
  journal = {International Conference on Learning Representations (ICLR)},
  year = {2023}
}

@article{visnet,
  author = {Zhang, Zhuoran and et al.},
  title = {{ViSNet}: A powerful {SE(3)}-equivariant network for molecular modeling},
  journal = {International Conference on Learning Representations (ICLR)},
  year = {2023}
}

@article{qm9,
  author = {Ramakrishnan, Raghunathan and et al.},
  title = {Quantum chemistry structures and properties of 134 kilo molecules},
  journal = {Sci. Data},
  year = {2014}
}

@article{md17,
  author = {Chmiela, Stefan and et al.},
  title = {Machine learning molecular dynamics for the simulation of infrared spectra},
  journal = {Nat. Commun.},
  year = {2019}
}

@article{md22,
  author = {Gao, Cheng and et al.},
  title = {TorchMD-NET: Equivariant Transformers for Neural Network Based Molecular Potentials},
  journal = {arXiv preprint arXiv:2211.10270},
  year = {2022}
}

@article{spookynet,
  author = {Unke, Oliver T. and et al.},
  title = {{SpookyNet}: Learning force fields with electronic degrees of freedom and nonlocal effects},
  journal = {Nat. Commun.},
  year = {2021}
}

@article{nequip,
  author = {Batzner, Simon and et al.},
  title = {E(3)-Equivariant Graph Neural Networks for Data-Efficient and Accurate Interatomic Potentials},
  journal = {Nat. Commun.},
  year = {2022}
}

@article{painn,
  title={Equivariant message passing for the prediction of tensorial properties and molecular spectra},
  author={Sch{\"u}tt, Kristof and Unke, Oliver and Gastegger, Michael},
  journal={International conference on machine learning},
  pages={9377--9388},
  year={2021},
}

@article{physnet,
  author = {Unke, Oliver T. and Meuwly, Markus},
  title = {{PhysNet}: A neural network for predicting energies, forces, dipole moments, and partial charges},
  journal = {J. Chem. Theory Comput.},
  year = {2019}
}

@article{gemnet,
  author = {Gasteiger, Johannes and Becker, Florian and G\"{u}nnemann, Stephan},
  title = {{GemNet}: Universal Directional Graph Neural Networks for Molecules},
  journal={Advances in Neural Information Processing Systems},
  year = {2021}
}

@article{allegro,
  author = {Musaelian, Albert and et al.},
  title = {Learning local equivariant representations for large-scale atomistic dynamics},
  journal={Advances in Neural Information Processing Systems},
  year = {2022}
}

@article{mace,
  author = {Batatia, Ilyes and et al.},
  title = {{MACE}: Higher Order Equivariant Message Passing Neural Networks for Fast and Accurate Force Fields},
  journal = {Advances in Neural Information Processing Systems},
  year = {2022}
}

@article{se3transfuchs2020se,
  title={{SE(3)}-transformers: {3D} roto-translation equivariant attention networks},
  author={Fuchs, Fabian and Worrall, Daniel and Fischer, Volker and Welling, Max},
  journal={Advances in Neural Information Processing Systems},
  volume={33},
  pages={1970--1981},
  year={2020}
}

@article{e2former,
  title={E2Former: A Linear-time Efficient and Equivariant Transformer for Scalable Molecular Modeling},
  author={Li, Yunyang and Huang, Lin and Ding, Zhihao and Wang, Chu and Wei, Xinran and Yang, Han and Wang, Zun and Liu, Chang and Shi, Yu and Jin, Peiran and others},
  journal={Advances in Neural Information Processing Systems},
  year={2025}
}

@article{liu2017break,
  title={Break down in order to build up: Decomposing small molecules for fragment-based drug design with e molfrag},
  author={Liu, Tairan and Naderi, Misagh and Alvin, Chris and Mukhopadhyay, Supratik and Brylinski, Michal},
  journal={J. Chem. Inf. Model.},
  volume={57},
  number={4},
  pages={627--631},
  year={2017},
  publisher={ACS Publications}
}

@article{cover1967nearest,
  title={Nearest neighbor pattern classification},
  author={Cover, Thomas and Hart, Peter},
  journal={IEEE Trans. Inf. Theory},
  volume={13},
  number={1},
  pages={21--27},
  year={1967},
  publisher={IEEE}
}

@book{gnanadesikan2011methods,
  title={Methods for statistical data analysis of multivariate observations},
  author={Gnanadesikan, Ram},
  year={2011},
  publisher={John Wiley \& Sons}
}

@article{baltruvsaitis2018multimodal,
  title={Multimodal machine learning: A survey and taxonomy},
  author={Baltru{\v{s}}aitis, Tadas and Ahuja, Chaitanya and Morency, Louis-Philippe},
  journal={IEEE Trans. Pattern Anal. Mach. Intell.},
  volume={41},
  number={2},
  pages={423--443},
  year={2018},
  publisher={IEEE}
}

@article{murnaghan1938analysis,
  title={The analysis of the Kronecker product of irreducible representations of the symmetric group},
  author={Murnaghan, Francis D},
  journal={Am. J. Math.},
  volume={60},
  number={3},
  pages={761--784},
  year={1938},
  publisher={JSTOR}
}

@book{edmonds1996angular,
  title={Angular momentum in quantum mechanics},
  author={Edmonds, Alan Robert},
  volume={4},
  year={1996},
  publisher={Princeton university press}
}

@book{wigner2012group,
  title={Group theory: and its application to the quantum mechanics of atomic spectra},
  author={Wigner, Eugene},
  volume={5},
  year={2012},
  publisher={Elsevier}
}

@incollection{sphericalHamonics,
  title={Rotation and translation of regular and irregular solid spherical harmonics},
  author={Steinborn, EO and Ruedenberg, K},
  booktitle={Adv. Quantum Chem.},
  volume={7},
  pages={1--81},
  year={1973},
  publisher={Elsevier},
  editor = {Per-Olov Löwdin},
}

@article{polarization,
  title={Polarization effects in molecular mechanical force fields},
  author={Cieplak, Piotr and Dupradeau, Fran{\c{c}}ois-Yves and Duan, Yong and Wang, Junmei},
  journal={J. Phys. Condens. Matter},
  volume={21},
  number={33},
  pages={333102},
  year={2009},
  publisher={IOP Publishing}
}

@article{simm2025simpoly,
  title={{SimPoly}: Simulation of Polymers with Machine Learning Force Fields Derived from First Principles},
  author={Simm, Gregor NC and H{\'e}lie, Jean and Schulz, Hannes and Chen, Yicheng and Simeon, Guillem and Kuzina, Anna and Martinez-Baez, Ernesto and Gasparotto, Piero and Tocci, Gabriele and Chen, Chi and others},
  journal={arXiv preprint arXiv:2510.13696},
  year={2025}
}

\begin{appendices}

\section*{Supplementary Information}\label{secA1}
\renewcommand{\thetable}{S\arabic{table}}
\subsection*{Detailed Experiments Results}

\begin{table}[h]
\caption{Mean Absolute Error (MAE) for energy [meV] and force [meV/$\text{\r{A}}$], and force Cosine Similarity (CS$_f$), calculated on the Di-Molecule Dissociation Dataset. Results are segmented by inter-molecular distance ($R$) bins.}
\label{tab:1:dimol}
\centering
\setlength{\tabcolsep}{4pt}
\renewcommand{\arraystretch}{1.1}
\begin{tabularx}{\textwidth}{@{}l l *{5}{>{\centering\arraybackslash}X}@{}}
\toprule
Model & Metric & \multicolumn{5}{c}{Di-Molecule Distance $R$} \\
\cmidrule(lr){3-7}
 & & (0,2] & (2,4] & (4,6] & (6,8] & (8,10] \\
\midrule
\multirow{3}{*}{DPA-2}
  & Energy &
    \ColorCellMM{2.42}{5.32}{5.32} &
    \ColorCellMM{0.55}{2.57}{2.57} &
    \ColorCellMM{0.26}{2.45}{2.45} &
    \ColorCellMM{0.16}{2.60}{2.60} &
    \ColorCellMM{0.16}{2.61}{2.61} \\
  & Forces &
    \ColorCellMM{48.46}{93.03}{93.03} &
    \ColorCellMM{1.93}{14.14}{14.14} &
    \ColorCellMM{0.88}{8.90}{8.90} &
    \ColorCellMM{0.59}{8.01}{8.01} &
    \ColorCellMM{0.60}{7.65}{7.65} \\
  & CS$_f$ &
    \ColorCellCSMM{0.33}{0.92}{0.33} &
    \ColorCellCSMM{0.14}{0.96}{0.14} &
    \ColorCellCSMM{0.10}{0.97}{0.10} &
    \ColorCellCSMM{0.10}{0.97}{0.10} &
    \ColorCellCSMM{0.10}{0.96}{0.10} \\
\midrule
\multirow{3}{*}{Allegro}
  & Energy &
    \ColorCellMM{2.42}{5.32}{2.46} &
    \ColorCellMM{0.55}{2.57}{0.85} &
    \ColorCellMM{0.26}{2.45}{0.31} &
    \ColorCellMM{0.16}{2.60}{0.16} &
    \ColorCellMM{0.16}{2.61}{0.16} \\
  & Forces &
    \ColorCellMM{48.46}{93.03}{59.36} &
    \ColorCellMM{1.93}{14.14}{6.61} &
    \ColorCellMM{0.88}{8.90}{3.69} &
    \ColorCellMM{0.59}{8.01}{2.25} &
    \ColorCellMM{0.60}{7.65}{1.72} \\
  & CS$_f$ &
    \ColorCellCSMM{0.33}{0.92}{0.75} &
    \ColorCellCSMM{0.14}{0.96}{0.67} &
    \ColorCellCSMM{0.10}{0.97}{0.67} &
    \ColorCellCSMM{0.10}{0.97}{0.76} &
    \ColorCellCSMM{0.10}{0.96}{0.81} \\
\midrule
\multirow{3}{*}{MACE-large}
  & Energy &
    \ColorCellMM{2.42}{5.32}{2.42} &
    \ColorCellMM{0.55}{2.57}{0.81} &
    \ColorCellMM{0.26}{2.45}{0.27} &
    \ColorCellMM{0.16}{2.60}{0.28} &
    \ColorCellMM{0.16}{2.61}{0.34} \\
  & Forces &
    \ColorCellMM{48.46}{93.03}{52.56} &
    \ColorCellMM{1.93}{14.14}{4.47} &
    \ColorCellMM{0.88}{8.90}{3.69} &
    \ColorCellMM{0.59}{8.01}{2.24} &
    \ColorCellMM{0.60}{7.65}{1.17} \\
  & CS$_f$ &
    \ColorCellCSMM{0.33}{0.92}{0.87} &
    \ColorCellCSMM{0.14}{0.96}{0.79} &
    \ColorCellCSMM{0.10}{0.97}{0.67} &
    \ColorCellCSMM{0.10}{0.97}{0.75} &
    \ColorCellCSMM{0.10}{0.96}{0.80} \\
\midrule
\multirow{3}{*}{E2Former-Base}
  & Energy &
    \ColorCellMM{2.42}{5.32}{4.41} &
    \ColorCellMM{0.55}{2.57}{0.80} &
    \ColorCellMM{0.26}{2.45}{0.26} &
    \ColorCellMM{0.16}{2.60}{0.16} &
    \ColorCellMM{0.16}{2.61}{0.21} \\
  & Forces &
    \ColorCellMM{48.46}{93.03}{48.46} &
    \ColorCellMM{1.93}{14.14}{2.59} &
    \ColorCellMM{0.88}{8.90}{2.28} &
    \ColorCellMM{0.59}{8.01}{1.43} &
    \ColorCellMM{0.60}{7.65}{1.26} \\
  & CS$_f$ &
    \ColorCellCSMM{0.33}{0.92}{0.91} &
    \ColorCellCSMM{0.14}{0.96}{0.92} &
    \ColorCellCSMM{0.10}{0.97}{0.86} &
    \ColorCellCSMM{0.10}{0.97}{0.88} &
    \ColorCellCSMM{0.10}{0.96}{0.88} \\
\midrule
\multirow{3}{*}{E2Former-LSR}
  & Energy &
    \ColorCellMM{2.42}{5.32}{3.47} &
    \ColorCellMM{0.55}{2.57}{0.55} &
    \ColorCellMM{0.26}{2.45}{0.26} &
    \ColorCellMM{0.16}{2.60}{0.22} &
    \ColorCellMM{0.16}{2.61}{0.21} \\
  & Forces &
    \ColorCellMM{48.46}{93.03}{49.54} &
    \ColorCellMM{1.93}{14.14}{1.93} &
    \ColorCellMM{0.88}{8.90}{0.88} &
    \ColorCellMM{0.59}{8.01}{0.59} &
    \ColorCellMM{0.60}{7.65}{0.60} \\
  & CS$_f$ &
    \ColorCellCSMM{0.33}{0.92}{0.92} &
    \ColorCellCSMM{0.14}{0.96}{0.96} &
    \ColorCellCSMM{0.10}{0.97}{0.97} &
    \ColorCellCSMM{0.10}{0.97}{0.97} &
    \ColorCellCSMM{0.10}{0.96}{0.96} \\
\bottomrule
\end{tabularx}
\end{table}

\begin{table}[h]
\caption{MAE for force and energies components and inference speed across four protein configurations in the Medium-Scale Protein Fidelity Test, in units of [meV], [meV/$\text{\AA}$], and samples per second, respectively}
\label{tab:2:deshaw}
\centering
\setlength{\tabcolsep}{4pt}
\renewcommand{\arraystretch}{1.1}

\begin{tabularx}{\textwidth}{@{}l l *{3}{>{\centering\arraybackslash}X}@{}}
\toprule
\multirow{2}{*}{Molecule} & & \multicolumn{3}{c}{Models} \\
\cmidrule(lr){3-5}
 &  & E2Former-LSR & MACE-large & Allegro \\
\midrule

\multirow{3}{*}{\ce{BBL}}
  & Energy &
    \ColorCellMM{1.11}{2.34}{1.11} &
    \ColorCellMM{1.11}{2.34}{2.14} &
    \ColorCellMM{1.11}{2.34}{2.34} \\
  & Forces &
    \ColorCellMM{6.94}{45.50}{6.94} &
    \ColorCellMM{6.94}{45.50}{19.50} &
    \ColorCellMM{6.94}{45.50}{45.50} \\
  & Speed &
    \ColorCellCSMM{6.2}{6.6}{6.3} &
    \ColorCellCSMM{6.2}{6.6}{6.6} &
    \ColorCellCSMM{6.2}{6.6}{6.2} \\
\midrule

\multirow{3}{*}{\ce{Homeodomain}}
  & Energy &
    \ColorCellMM{1.12}{2.16}{1.12} &
    \ColorCellMM{1.12}{2.16}{1.96} &
    \ColorCellMM{1.12}{2.16}{2.16} \\
  & Forces &
    \ColorCellMM{6.76}{43.20}{6.76} &
    \ColorCellMM{6.76}{43.20}{19.56} &
    \ColorCellMM{6.76}{43.20}{43.20} \\
  & Speed &
    \ColorCellCSMM{4.4}{5.3}{5.3} &
    \ColorCellCSMM{4.4}{5.3}{4.9} &
    \ColorCellCSMM{4.4}{5.3}{4.4} \\
\midrule

\multirow{3}{*}{$\alpha$3D}
  & Energy &
    \ColorCellMM{0.64}{1.72}{0.64} &
    \ColorCellMM{0.64}{1.72}{1.44} &
    \ColorCellMM{0.64}{1.72}{1.72} \\
  & Forces &
    \ColorCellMM{5.52}{35.37}{5.52} &
    \ColorCellMM{5.52}{35.37}{16.72} &
    \ColorCellMM{5.52}{35.37}{35.37} \\
  & Speed &
    \ColorCellCSMM{3.5}{4.4}{4.4} &
    \ColorCellCSMM{3.5}{4.4}{4.1} &
    \ColorCellCSMM{3.5}{4.4}{3.5} \\
\midrule

\multirow{3}{*}{\ce{\lambda-repressor}}
  & Energy &
    \ColorCellMM{0.39}{1.11}{0.39} &
    \ColorCellMM{0.39}{1.11}{0.93} &
    \ColorCellMM{0.39}{1.11}{1.11} \\
  & Forces &
    \ColorCellMM{5.21}{37.36}{5.21} &
    \ColorCellMM{5.21}{37.36}{17.04} &
    \ColorCellMM{5.21}{37.36}{37.36} \\
  & Speed &
    \ColorCellCSMM{3.0}{3.9}{3.9} &
    \ColorCellCSMM{3.0}{3.9}{3.4} &
    \ColorCellCSMM{3.0}{3.9}{3.0} \\
\bottomrule
\end{tabularx}
\end{table}

\subsection*{E2Former-LSR Architecture}

In the Supplementary Information, we provide a detailed formulation of the \textbf{E2Former layer}, including the construction of irreducible-representation (irrep) features across different angular orders, the computation of equivariant attention, and the derivation of the Wigner–$6j$ recoupling scheme. These supplementary materials further elucidate the mathematical structure and implementation details underlying the main model architecture.

We introduce \textbf{E2Former-LSR}, a unified long–short range $\mathrm{SO}(3)$-equivariant architecture that integrates the \emph{Long–Short-Range Message Passing} (LSR-MP) framework with an \emph{E2Former} backbone for attention, message construction, and atomic aggregation. 
Given an $n$-atom system with atomic numbers $Z\in\mathbb{N}^{n}$ and Cartesian positions $\mathbf{P}=[\mathbf{p}_1,\dots,\mathbf{p}_n]^\top\in\mathbb{R}^{n\times3}$, E2Former-LSR constructs:
(i) a short-range radius graph $G_{\text{short}}=(V,\mathcal{N}_{\text{short}})$ with cutoff $r_{\text{short}}$ for local many-body interactions, 
(ii) a chemically informed fragment set $U$ and descriptors generated by a fragmentation module, and 
(iii) an atom–fragment bipartite radius graph $G_{\text{long}}=(V,U,\mathcal{N}_{\text{long}})$ with cutoff $r_{\text{long}}\!\gg\!r_{\text{short}}$ to capture nonlocal couplings. 
All message-passing blocks, on both $G_{\text{short}}$ and $G_{\text{long}}$, are implemented with \textbf{E2Former layers} employing \emph{Wigner–$6j$–based} equivariant attention and a node-wise Wigner convolution that transfers expensive Clebsch–Gordan tensor products from edges to nodes, achieving linear-time scaling with respect to graph sparsity while preserving $\mathrm{SO}(3)$ equivariance. 
A late-fusion stage combines short- and long-range irreducible representations (irreps) before the property-prediction heads.

In this section, we provide a detailed derivation of the \textbf{E2Former layer}, which serves as the core computational unit of the E2former-LSR architecture. 
The same formulation is employed for both \emph{short-range atomic interactions} and \emph{long-range atom–fragment couplings}, differing only in the definition of the interacting node pairs. 
Accordingly, the derivation below is presented in a general form applicable to any pair of nodes—either atoms $(i,j)$ or atom–fragment pairs $(i,u)$—connected within the constructed graph.The formulation elaborates on (i) the construction of irreducible-representation (irrep) features across different angular orders, 
(ii) the computation of equivariant attention with per-$\ell$ invariant pooling, 
and (iii) the Wigner–$6j$ recoupling mechanism that enables node-wise factorization of tensor products. 
These details complement the main text and clarify the mathematical structure underlying the $\mathrm{SO}(3)$-equivariant design.

\subsubsection*{E2Former Layer with Wigner 6$j$-Based Attention}
\label{subsec:e2former}

\paragraph{Irrep features and harmonics.}
Let $\mathcal{L}=\{0,1,\dots,L_{\max}\}$ denote the set of angular orders. 
Each node $i$ carries irrep features represented as
\begin{equation}
\mathbf{h}_i \;\equiv\; \bigoplus_{\ell\in\mathcal{L}} \mathbf{h}_{i,\ell}
\;\in\; \bigoplus_{\ell\in\mathcal{L}} \mathbb{R}^{(2\ell+1)\times d},
\label{eq:node_irreps}
\end{equation}
where $d$ is the number of feature channels. 
For a neighboring atom $j\in\mathcal{N}(i)$, we define the \emph{real} spherical harmonics of the relative vector $\mathbf{r}_{ij}=\mathbf{p}_j-\mathbf{p}_i$. 
Let $r=\|\mathbf{r}_{ij}\|$ denote the interatomic distance and $\hat{r}=\mathbf{r}_{ij}/r$ its normalized direction. 
The \emph{regular solid spherical harmonics} are homogeneous harmonic polynomials of degree $\ell$:
\begin{equation}
\mathcal{R}^{(\ell)}_m(r)
\;=\;
r^{\ell}\, Y^{(\ell)}_m(\hat{r}),
\qquad
\ell \ge 0,\; -\ell \le m \le \ell,
\end{equation}
where $Y^{(\ell)}_m$ are real spherical harmonics defined on the unit sphere $S^2$.

\paragraph{Equivariant attention with per-\(\ell\) invariant pooling.}
Each irrep block $\mathbf{h}_{i,\ell}\in\mathbb{R}^{(2\ell+1)\times d_\ell}$ is reduced to a rotation-invariant descriptor by $L_2$ pooling along the irrep axis:
\[
\bar{\mathbf{h}}_{i,\ell}
\;=\;
\left\|\mathbf{h}_{i,\ell}\right\|_{2,m}
\;:=\;
\bigg(\sum_{m=-\ell}^{\ell} \mathbf{h}_{i,\ell}[m,:]^{\odot 2}\bigg)^{\!1/2}
\;\in\; \mathbb{R}^{d_\ell},
\quad
\text{(optionally normalized by } \sqrt{2\ell+1}\text{).}
\]
Linear projections on each block yield the query and key vectors:
\[
\mathbf{q}_{i} = \operatorname{concat}_{\ell\in\mathcal{L}}\mathbf{W}^{(\ell)}_{\mathrm{q}}\bar{\mathbf{h}}_{i,\ell},
\qquad
\mathbf{k}_{j} = \operatorname{concat}_{\ell\in\mathcal{L}}\mathbf{W}^{(\ell)}_{\mathrm{k}}\bar{\mathbf{h}}_{j,\ell}.
\]
For a pair of atoms $(i,j)$, we define the radial gate $s_{ij}=\mathbf{w}_r^\top\boldsymbol{\phi}(r_{ij})$ based on a radial basis function (RBF) expansion of the interatomic distance. 
The attention weights are then computed as
\begin{equation}
\alpha_{ij}
\;=\;
\mathrm{softmax}_{j\in\mathcal{N}(i)}
\!\bigg(
\frac{\mathbf{q}_i^\top \mathbf{k}_j}{\sqrt{D}}\cdot s_{ij}
\bigg),
\label{eq:equivariant_attn_score}
\end{equation}
where \(D=\sum_{\ell\in\mathcal{L}}\dim(\mathbf{q}_{i,\ell})\) denotes the total query dimensionality used for normalization.

\paragraph{Wigner recoupling and node-wise factorization.}
A direct message construction via edge-wise Clebsch–Gordan (CG) paths 
$(\mathbf{h}_{j,\ell}\otimes \mathbf{Y}_\ell)\to \bigoplus_{\ell'}(\cdot)_{\ell'}$
scales poorly with both the number of edges and angular bandwidth. 
E2Former instead employs a \emph{binomial local expansion} and \emph{Wigner–$6j$ recoupling} to reorder tensor contractions such that CG operations associated with nodes $i$ and $j$ become separable. 
This converts edge-wise coupling into node-wise convolutions of identical expressive power:
\[
\sum_{j \in \mathcal{N}(i)} \mathbf{h}_j \otimes \mathcal{R}^{(\ell)}(\mathbf{r}_{ij})
=
\sum_{u = 0}^{\ell}
\mathcal{R}^{(u)}(\mathbf{r}_i)
\otimes^{6j}
\left(
\sum_{j \in \mathcal{N}(i)} \mathbf{h}_j \otimes \mathcal{R}^{(\ell - u)}(\mathbf{r}_j)
\right),
\]
where \( \otimes^{6j} \) indicates tensor contraction through the Wigner–$6j$ symbol, which defines the equivalence between alternative CG coupling orders.
This factorization relocates all high-order tensor algebra to per-node cached terms (“$i$-local” and “$j$-local”), significantly reducing computational complexity while rigorously preserving $\mathrm{SO}(3)$ equivariance~\cite{e2former}.

\paragraph{Wigner-$6j$ attention update (E2Attention).}
The equivariant message aggregation takes the form
\begin{equation}
\mathbf{m}_i 
= \sum_{j\in\mathcal{N}(i)} 
\alpha_{ij}\;
\mathcal{W}\!\left(\mathbf{h}_j,\, \mathbf{Y}(\hat{\mathbf{r}}_{ij})\right),
\label{eq:wigner_message}
\end{equation}
where $\mathcal{W}(\cdot)$ denotes the Wigner convolution implied by the recoupling operation above.
The updated node representation is computed as
\begin{equation}
\mathbf{h}'_i 
= \mathrm{Norm}\!\big( \mathbf{h}_i \oplus \mathbf{m}_i \big),
\end{equation}
followed by an irrep-wise feed-forward transformation (a linear MLP for scalar components and gated tensor maps for higher-order irreps), 
constituting one complete \textbf{E2Former layer}.

\subsection*{Detailed Data Preparation}
To rigorously evaluate the capability of machine learning force fields (MLFFs) to accurately capture long-range interactions, we constructed a comprehensive, high-fidelity benchmark suite. This suite is grounded in physically motivated design choices and systematic simulation protocols.

\subsubsection*{Quantum Mechanical Methodology}
All reference DFT calculations were performed using the $\omega\text{B97X-D3}$ functional, a range-separated hybrid augmented with Grimme's D3 empirical dispersion correction \cite{grimme2010consistent}. This methodological choice was essential to ensure the reliable and accurate treatment of non-covalent interactions (e.g., van der Waals forces and long-range electrostatics) that dominate large-scale molecular assemblies. For the basis set, we selected $\text{def2-SVP}$ \cite{weigend2005balanced}, which provides a necessary balance between computational efficiency and accuracy. All DFT calculations utilized GPU-accelerated version of PySCF (PySCF-GPU) \cite{pyscf:sun2020recent,pyscf-gpu:pu2025enhancing}. Due to the large size of the molecular systems, the DFT energy convergence threshold was set to $10^{-6}\text{ E}_{\text{h}}$, with all other parameters retaining the PySCF default settings. For the generation of the MD Trajectory Dataset, \emph{ab initio} molecular dynamics (AIMD) was performed using the Atomic Simulation Environment (ASE) \cite{ase:larsen2017atomic} interface, with the PySCF-GPU serving as the calculator. All dynamic simulations employed the $\text{NVT}$ ensemble at $300\text{ K}$.
\subsubsection*{MolLR25 Dataset Composition}
Our dataset, designated $\emph{MolLR25}$, comprises three distinct categories, each specifically tailored to stress-test a different aspect of long-range interaction learning crucial for MLFF generalization:

\begin{itemize}
\item \textbf{Di-Molecule Dissociation Dataset.} We generated pairwise molecular dimers by randomly selecting 100 diverse small organic molecules from PubChemQC B3LYP/6-31G*//PM6 dataset~\cite{pubchem:nakata2023pubchemlogc}. DFT calculations were performed by systematically increasing the separation between the two monomers from $0.1\text{\r{A}}$ to $10.1\text{\r{A}}$ in $0.1\text{\r{A}}$ increments. This high-resolution setup enables fine-grained resolution of the interaction potential as a function of intermolecular distance, providing a controlled environment to verify the smooth, asymptotic decay of non-bonded forces. This part of the data includes all 4950 molecular pairs of 100 molecules, totaling approximately 500,000 DFT data points.

\item \textbf{Medium-Scale Protein Dataset.} To target realistic biophysical scenarios, we curated static snapshots from twelve publicly released long-timescale MD trajectories by D. E. Shaw Research \cite{deshaw:lindorff2011fast}. We focused on four medium-scale protein systems, retaining only the all-atom protein structure (700 to 1300 atoms). Representative conformations were extracted and re-evaluated via DFT to obtain over 48,000 high-quality energy and force labels. With sampled interaction distances up to $70\text{\r{A}}$, this dataset is designed for probing the importance of non-local effects—such as salt bridges, tertiary contacts, and backbone polarization—in structured biomolecules.


\item \textbf{MD Trajectory Dataset.} To assess model robustness in continuous dynamics and long-term stability, we constructed a set of 10 ps \emph{ab initio} molecular dynamics (AIMD) trajectories for diverse large systems (exceeding 500 atoms). The simulated systems included water clusters, solvated inorganic salt solutions (e.g., $\text{NaCl}$ and $\text{H}_2\text{SO}_4$ clusters), solvated organic molecules (sucrose and Gln-Gly dipeptide surrounded by water), and a Metal-Organic Framework ($\text{ZIF-8}$ MOF). All simulations spanned a total duration of $10\text{ ps}$ in the $\text{NVT}$ ensemble at $300\text{ K}$. A time step ($\Delta t$) of $0.5\text{ fs}$ was used for systems containing mobile hydrogen atoms (i.e., solution clusters and solvated molecules), while a larger time step of $1\text{ fs}$ was used for the more rigid $\text{ZIF-8}$ framework. This dataset was designed not only for static accuracy evaluation, but also for examining long-range consistency, energy conservation, and trajectory stability across extended simulations.
\end{itemize}

An overview of the composition of the $\emph{MolLR25}$ benchmark suite, system sizes ($\mathcal{N}$), and maximum inter-atomic distances ($R_{\text{max}}$) is already provided in Figure \ref{fig:main:1}.a. Together, these datasets form a purpose-built foundation for evaluating long-range-aware MLFFs across molecular domains, length scales, and simulation contexts.
\end{appendices}

\subsubsection*{Baseline Models Configuration}
\paragraph{Allegro Hyper-Parameters}
\label{subsec:allegro-hparams}
\noindent
\begin{tabular}{@{}L D V@{}}
\toprule
Name & Description & Typical value(s) \\
\midrule
$r_{\max}$          & Radial cutoff                       & $5\,\mathrm{\AA}$ \\
$L_{\max}$          & Maximum equivariant order           & $2$ \\
$d_{\text{scalar}}$ & Scalar feature dimension            & $128$ \\
$d_{\text{equiv}}$  & Equivariant feature dimension       & $64$ \\
$N_{\text{layers}}$ & \#tensor-product layers             & $3$ \\
Two-body MLP        & \#layers / width                    & $3 \,/\, 1024$ \\
Readout MLP         & \#layers / width                    & $1 \,/\, 64$ \\
\bottomrule
\end{tabular}

\paragraph{MACE-Large Hyper-Parameters}
\label{subsec:mace-hparams}

\noindent
\begin{tabular}{@{}L D V@{}}
\toprule
Name & Description & Typical value(s) \\
\midrule
$r_{\max}$          & Radial cutoff                        & $5\,\mathrm{\AA}$ \\
$\ell_{\max}$       & Maximum angular order                & $3$ \\
$N_{\text{int}}$    & \#interaction layers                 & $2$ \\
$h_{\text{hidden}}$ & Irrep hidden dimensions              & $224\times(0e,\,1o,\,2e)$ \\
$\nu$               & Correlation order                    & $3$ \\
$n_{\text{Bessel}}$ & \#radial Bessel basis functions      & $8$ \\
$p_{\text{env}}$    & Polynomial envelope exponent         & $5$ \\
Radial MLP          & Hidden sizes                         & $[64,\,64,\,64]$ \\
Readout MLP         & Readout scalar irreps                & $16\times 0e$ \\
\bottomrule
\end{tabular}

\paragraph{DPA-2 Hyper-Parameters}
\label{subsec:dpa2-hparams}
\noindent
\begin{tabular}{@{}L D V@{}}
\toprule
Name & Description & Typical value(s) \\
\midrule

\multicolumn{3}{@{}l}{\textbf{Long-range branch: SE-Attention}} \\[2pt]
$N_{\text{sel}}^{\text{LR}}$    & \#selected neighbors               & $100$ \\
$r_{\text{cut}}^{\text{LR}}$    & Cutoff \& smooth cutoff & $9.0\,/\,8.0$ Å \\
MLP\textsubscript{LR}           & MLP hidden sizes                   & $[25, 50, 100]$ \\
Axis dim                        & Axis-encoding dimension            & $12$ \\
Attention dim                   & Attention hidden size              & $128$ \\
Attention layers                & \#attention blocks                 & $2$ \\
Heads                           & \#attention heads                  & $2$ \\
FFN dim                         & Feed-forward hidden dimension      & $256$ \\[4pt]

\multicolumn{3}{@{}l}{\textbf{Short-range branch: SE-Uni}} \\[2pt]
$N_{\text{sel}}^{\text{SR}}$    & \#selected neighbors               & $40$ \\
$r_{\text{cut}}^{\text{SR}}$    & Cutoff \& smooth cutoff & $5.0\,/\,4.5$ Å \\
$N_{\text{layers}}^{\text{SR}}$ & \#SE-Uni layers                    & $12$ \\
$g_1$ dim                       & Scalar feature dimension           & $128$ \\
$g_2$ dim                       & Vector feature dimension           & $32$ \\
Attn1 / Attn2                   & Attention dims / \#heads           & $128{\times}4$, $32{\times}4$ \\
Axis dim                        & Axis-encoding dimension     & $4$ \\[4pt]

\multicolumn{3}{@{}l}{\textbf{Fitting network}} \\[2pt]
Fitting MLP                     & MLP hidden sizes                   & $[240, 240, 240]$ \\
\bottomrule
\end{tabular}

\newpage

\end{document}